\documentclass[preprint,nofootinbib]{revtex4}

\usepackage{graphicx}  
\usepackage{dcolumn}   
\usepackage{bm}        
\usepackage{amssymb}
\usepackage{diagbox}
\usepackage{slashbox}
\usepackage{slashbox}
\usepackage{hyperref}
\usepackage{natbib}
\usepackage{pifont}
\usepackage{multirow}
\usepackage{amsmath}
\usepackage{cases}
\usepackage{placeins}
\usepackage{textcomp}
\UseRawInputEncoding

\usepackage{blindtext}
\usepackage{url}
\usepackage{graphicx}
\usepackage{url}
\usepackage{tikz}
\usepackage[bf]{caption}
\usepackage{amsmath}
\usepackage{cases} 
\usepackage{multirow}

\usepackage{epsfig}
\usepackage{makecell}
\usepackage{amsmath}
\usepackage{amsfonts}
\usepackage{amssymb}
\usepackage{mathtools}
\usepackage{url}
\usepackage{subfigure}
\usepackage{hhline}
\usepackage{color}
\usepackage{array,multirow}
\usepackage{bm}
\usepackage{makecell}
\usepackage{epsfig}
\usepackage{amsmath}
\usepackage{amsfonts}
\usepackage{amssymb}
\usepackage{url}
\usepackage{hyperref}
\usepackage{subfigure}
\usepackage{hhline}
\usepackage{color}
\usepackage{bm}
\usepackage{cancel}
\usepackage{mathtools}

\newcommand{\beqa}{\begin{eqnarray}}
\newcommand{\eeqa}{\end{eqnarray}}
\newcommand{\beq}{\begin{equation}}
\newcommand{\eeq}{\end{equation}}

\newcommand{\nn}{\nonumber}
\newcommand{\bmt}{\begin{pmatrix}}
\newcommand{\emt}{\end{pmatrix}}
\usepackage[toc,page]{appendix}
\usepackage{comment}
\newcommand{\be}{\begin{equation}}
\newcommand{\ee}{\end{equation}}
\newcommand{\bea}{\begin{eqnarray}}
\newcommand{\eea}{\end{eqnarray}}

\hypersetup{
    colorlinks=true,
    citecolor=red,
    linkcolor=blue,
    filecolor=green,      
    urlcolor=magenta,
}

\begin{document}
\title{\bf Exploring the lepton flavor violating decay  modes $b \to s \mu ^{\pm} \tau ^{\mp}$  in SMEFT approach}

\author{Dhiren Panda}
\email{pandadhiren530@gmail.com}
\author{Manas Kumar Mohapatra}
\email{manasmohapatra12@gmail.com}

\author{Rukmani Mohanta}
\email{rmsp@uohyd.ac.in}
\affiliation{School of Physics,  University of Hyderabad, Hyderabad-500046,  India}

\begin{abstract}
We perform an analysis of the consequences of various new physics operators on the lepton flavor violating (LFV)  decay modes mediated through $b \to s \ell _1 \ell _2$ transitions. We scrutinize the imprints of the (pseudo)scalar and axial(vector) operators on the exclusive LFV decay channels $ B_{(s)} \rightarrow (\phi, K^{*}, K_{2}^{*})\ell_{1}\ell_{2}$  and  $\Lambda_{b}\rightarrow \Lambda \ell_{1}\ell_{2}$, where $\ell_{1},  \ell_{2}$ represent $\mu$ or $\tau$. The new physics parameters are constrained by using the upper limits of the branching fractions of the $B_s \to \tau \mu$ and $B \to K \tau \mu$ processes, assuming the new physics couplings to be real. We then explore the key observables such as the branching fractions, the forward-backward asymmetries, and the longitudinal polarisation fractions of the $B \to (K^*, \phi, K_2^*) \tau ^{\pm} \mu ^{\mp}$ decays. 
In addition, we also investigate the impact of the new physics couplings on the baryonic $\Lambda _b \to \Lambda \tau ^{\pm} \mu ^{\mp}$ decay channels mediated by the $b \to s$ quark level transition. With the experimental prospects at LHCb upgrade and Belle II, we also predict the upper limits of the above-discussed observables, which could intrigue the new physics search in these channels.
\end{abstract}

\maketitle
\section{Introduction}
The Standard Model (SM) delineates the elementary particles and their interactions and provides a comprehensive framework that firmly establishes a wide range of natural phenomena occurring at energies below the electroweak scale. Despite its remarkable achievements, the pursuit of physics beyond the Standard Model (BSM) remains crucial. This quest addresses numerous experimental anomalies and theoretical puzzles, including the prevalence of matter over antimatter in the universe, the enigma of dark matter and dark energy, as well as the hierarchy and flavor problems, among others.
The potential existence of physics beyond the Standard Model arises from two distinct avenues. The first involves indirect effects of new physics (NP) associated with the presence of heavy new particles, which can modify the Wilson coefficients of the interaction Hamiltonian within the Standard Model framework. Alternatively, direct approaches entail detecting new particles through ongoing and upcoming collider experiments. Among various searches for physics beyond the SM, the $B$ meson decays have been considered as one of the potential avenues for BSM search in the indirect approach for many years. In this respect, the flavor-changing neutral current transition (FCNC) mediated decays, especially the $(b \to s/d)$ transitions are more captivating for understanding the nature of rare $b$ quark decays. The LHC at CERN, particularly LHCb \cite{LHCb:2022qnv, LHCb:2022vje} collaboration recently confirmed the lepton flavor universality
(LFU) violating ratio $R_{K^{(*)}}=\mathcal{B} (B \to K^{(*)} \mu \mu)/\mathcal{B} (B \to K^{(*)} e e)$ to be  consistent with the SM, which are of the order of unity. On the other hand, there exist various other observables such as the form factor independent (FFI) observable $P_5^{\prime}$ in $B \to K^* \mu^- \mu^+$ process and the branching ratios of several decay channels display few sigma deviations from the SM values. The LHCb \cite{LHCb:2013ghj, LHCb:2015svh}  and ATLAS \cite{ATLAS:2018gqc} collaborations reported a 3.3$\sigma$ difference from the SM prediction in the measurement of the celebrated $P_5^{\prime}$ observable. The other LFU violating observables, respecting the $b \to s$ quark level transitions, $R_{K_S^0}$ and $R_{K^{*+}}$ \cite{LHCb:2021lvy} also show deviations at the level of $1.4 \sigma$ and $1.5 \sigma$ compared to the SM predictions, respectively. On the other hand, the branching fraction of the rare $B_s \to \phi \mu \mu$ decay mode reveals $3.3 \sigma$ \cite{LHCb:2021zwz} away from the SM value in $q^2 \in[1.1,6.0]~ \rm GeV^2$. 

Unlike the lepton flavor-conserving decays, the LFV transitions indicate a clean probe of NP as they are forbidden in the SM. Various  LFV decays in the charged lepton sector i.e.,  $\ell _i \to \ell _j \gamma$, $\ell _i \to \ell _j \ell _k \bar{\ell} _k$ as well as in $B$ meson sector, mediated through $b \to s \ell _i \ell _j$ transitions, have been extensively studied in the literature \cite{Lee:2015qra, Altmannshofer:2015mqa, Crivellin:2015mga, Alonso:2015sja, Sahoo:2015pzk, Sahoo:2015wya, Mohapatra:2023hhf, Ali:2023kua}. The LFV $B$ meson decays provide an ideal platform for NP search, particularly the results from  LHCb and Belle II experiments can be used as a guiding principle in this direction. However, so far we have only the experimental upper limits for these decay modes. The LHCb experiment reported the upper bound on the branching ratio of the purely leptonic $B_s \to e^{\pm} \mu ^{\mp}$ decay channel, which is found to be $\mathcal{B} (B_s \to e^{\pm} \mu ^{\mp}) < 6.3 \times 10^{-9}$ at 90\%  C.L. In addition, an upper limit of $4.2 \times 10^{-5}$ in the branching fraction of $B_s \to \tau \mu$ has also been reported by the LHCb experiment \cite{LHCb:2019ujz}. The LFV searches in semileptonic $B$ decays, on the other hand, include well-known decays like  $B \to K \ell _1 \ell _2$ processes, where $\ell _1 $ or $\ell _2 = e, \mu, \tau$. Based on Run I data with integrated luminosity 9 ${\rm fb}^{-1}$, an exclusion limit on $\mathcal{B} (B \to K \mu ^- e^+) < 7 \times 10^{-9}$ is reported by LHCb  collaboration \cite{LHCb:2019bix}. Similar searches  were also performed for $B \to V \ell _1 \ell _2$ ($V = K^*, \phi$) processes. The current limit on the branching ratios of $B \to K^* \mu e$ and $B \to \phi \mu e$, set by LHCb, are $6.8 \times 10^{-9}$ and $10.1 \times 10^{-9}$, respectively. Despite the experimental difficulties due to missing energy during reconstruction from the environment of $\tau$ lepton present in the final state of the $B$ meson decays, Belle \cite{Belle:2022pcr} experiment put an upper limit on the branching ratio of $B^+ \to K^+ \mu ^{\pm} \tau ^{\mp}$ decay mode as $3.9 \times 10^{-5}$. An analysis is performed using the LHCb data \cite{LHCb:2022wrs} on the $B \to K^* \mu ^{\pm} \tau ^{\mp}$ process. However, no signal is observed, instead, an upper limit on the branching fraction of $B \to K^* \tau ^+ \mu ^-$ and $B \to K^* \tau ^- \mu ^+$  are set to be $1.0 \times 10^{-5}$ and $8.2 \times 10^{-6}$ \cite{LHCb:2022wrs} at the 90$\%$ confidence level.

In this work, we intend to explore the effect of the SMEFT (Standard Model Effective Field Theory) formalism on the exclusive semileptonic $b \to s \tau \mu$ decays. Given the current and future experimental prospects, we mainly focus on the $B \to (K^*, \phi, K_2^*) \tau ^{\pm} \mu ^{\mp}$ decays. In the $B \to K_2^* \tau \mu$ decay, the tensor meson $K_2^*$ includes additional polarization states compared to the $K^*$, leading to new kinematic observables sensitive to novel insights into new physics. Additionally, the $b \to s \tau \mu$ mediated modes also offer an unique opportunity to probe new helicity structures, complementing the recently measured $B \to K^{0*} \mu \tau $ channel \cite{LHCb:2022wrs}.  Furthermore, Belle \cite{Belle:2002ekk} and BaBar \cite{BaBar:2003aji} have observed the radiative decay $B \to K_2^* \gamma$, with a branching ratio comparable to $B \to K^* \gamma$. This suggests that $K_2^*$ states could play a significant role in rare $B$-meson decays, further motivating the study of $B \to K_2^* \tau \mu$ as a potential avenue for new physics. A recent study investigating the lepton-flavor violating decay $B_s \to \phi \mu \tau$ \cite{LHCb:2024wve}, utilizing proton-proton collision data at 7, 8, and 13 TeV center-of-mass energies, gathered by the LHCb detector, with a total integrated luminosity of $9~ \text{fb}^{-1}$. However, the analysis could not achieve any significant signal, leading to an upper limit on the branching fraction of $\mathcal{B}(B_s \to \phi \mu \tau) < 1.0 \times 10^{-5}$ with 90$\%$ confidence level.
 
In addition, we also investigate the impact of the new physics couplings on the baryonic $\Lambda _b \to \Lambda \tau ^{\pm} \mu ^{\mp}$ decay channels mediated by $b \to s$ quark level transition. Our fit anatomy includes the upper limit of the branching ratios of the leptonic $B_s \to \tau \mu$ and semileptonic $B \to K \tau \mu$ processes. Using the constrained values of the NP couplings, we probe the prominent observables such as the branching fraction, the forward-backward asymmetry, and the longitudinal polarisation fraction of the above decay modes in the presence of various SMEFT operators. 

The paper is organized in the following manner. In section \ref{subsec:theory}, we recapitulate the theoretical framework of the EFT framework of $ b \rightarrow s \ell_{1} \ell_{2}$  transition. We also discuss on various observable associated with the $B_{(s)}\to (K^*,\phi,K_2^{*}) \ell_1 \ell_2$ and $\Lambda_{b}\to \Lambda \ell_1 \ell_2$ processes. In section III, we constrain the new physics parameter space in the presence of various SMEFT operators. We interpret the outcome followed by numerical analysis of the $B \to (K^*, \phi, K_2^*) \tau ^{\pm} \mu ^{\mp}$ and the baryonic $\Lambda _b \to \Lambda \tau ^{\pm} \mu ^{\mp}$ decays in the presence of new physics coefficients. Finally, we conclude our work in section IV.
\section{Theoretical Framework}
\label{subsec:theory}
The most generic effective Hamiltonian describing the $b \to s\ell_{1}^-\ell_{2}^+$ decay  is given as \cite{Das:2019omf}
 \begin{equation}\label{eq:234}
     \begin{split}
         \mathcal H_{eff} =-\frac{4G_{F}}{\sqrt{2}}V_{tb}V^{*}_{ts}\frac{\alpha_{em}}{4\pi}\sum_{i=9, 10,S,P}\big(C_{i}^{\ell_1\ell_2}(\mu)\mathcal{O}_{i}^{\ell_1\ell_2}(\mu)+C_{i}^{'\ell_1\ell_2}(\mu)\mathcal{O'}_{i}^{\ell_1\ell_2}(\mu)\big),  
\end{split}
 \end{equation}
where $G_{F}$ and $V_{tb}V_{ts}^*$ are the Fermi coupling constant and CKM matrix elements respectively. 
The relevant operators for the process are expressed as follows,
 \begin{eqnarray}\label{eq:opbasis}
\begin{split}
&\mathcal{O}^{(\prime)}_9 = \big[\bar{s}\gamma^\mu P_{L(R)}b \big]\big[\ell_2\gamma_\mu\ell_1 \big]\, 
 ,\quad \mathcal{O}^{(\prime)}_{10} = \big[\bar{s}\gamma^\mu P_{L(R)}b \big]\big[\ell_2\gamma_\mu\gamma_5\ell_1 \big]\;,\\
&\mathcal{O}_S^{(\prime)} = \big[\bar{s}P_{R(L)}b \big]\big[\ell_2\ell_1 \big]\, ,\quad \quad \quad \mathcal{O}_P^{(\prime)} = \big[\bar{s}P_{R(L)}b \big]\big[\ell_2\gamma_5\ell_1 \big]
\,.
\end{split}
\end{eqnarray}

    Here, the operators $\mathcal{O}_{9,10, S, P}$ represent the vector, axial-vector, scalar, and pseudoscalar operators, respectively. The primed operators $\mathcal{O}_i^{\prime}$ can be obtained by flipping the chirality of the former operators $\mathcal{O}_i$. The $C^{(\prime)}_{9, 10, S, P}$  are the Wilson coefficients that have zero value in the SM and can have a non-zero value in various new physics scenarios. In SM, the leptons $\ell_1,\ell_2$ correspond to the same flavor which is usually considered as $\ell$. 
    
 \subsection*{The decay observables of $b \to s \ell_1 \ell_2$ transitions}
\subsubsection{\textbf{Exclusive $\Lambda_b \rightarrow \Lambda \ell_1\ell_2 $ decay channel}}
To illustrate the kinematics of the $\Lambda_b \rightarrow \Lambda \ell_1\ell_2 $ decay mode, we assume the baryon $\Lambda_b$ is at rest, while the final state particles $\Lambda$ and the dilepton pair travel along the positive and negative $z$-axis, respectively. The momenta assigned to the particles $\Lambda_{b}$, $\Lambda$, $\ell_1$, and $\ell_2$ are represented by the symbols $p$, $k$, $q_1$, and $q_2$ respectively. The spin of the baryon  $ \Lambda_{b}$ ($\Lambda$)  on to the z-axis in the rest frame is denoted as $s_p (s_k)$. The decay amplitude of the exclusive $\Lambda_b \rightarrow \Lambda \ell_1\ell_2 $ process can be written as \cite{Das:2019omf},
\begin{equation}
\begin{split}
    \mathcal{M}^{\lambda_1,\lambda_2}(s_p,s_k)=-\frac{V_{tb}V_{ts}^*}{2v^2}\frac{\alpha_{em}}{4\pi}\sum_{i=L,R}\Big[\sum_{\lambda}\eta_{\lambda}H_{VA,\lambda}^{i,s_p,s_k}L_{i,\lambda}^{\lambda_2,\lambda_1}+H_{SP}^{i,s_p,s_k}L_{i}^{\lambda_2,\lambda_1}\Big].
\end{split}
\end{equation}

In the dilepton rest frame, $q^{\mu }$ represents the four-momentum of the dilepton pair, and $\theta_{\ell}$ denotes the angle between the $\ell_1$ and $z$-axis of the dilepton rest frame.
Here the hadronic helicity amplitude $H_{VA (SP),\lambda}^{i,s_p,s_k}$ correspond to the vector-axial vector (scalar-pseudoscalar) operators whereas the  $L_{i,\lambda}^{\lambda_2,\lambda_1}$, and $L_{i}^{\lambda_2,\lambda_1}$ are the leptonic helicity amplitudes.\\

For the detailed expression of the lepton helicity amplitude, we refer to Ref. \cite{Das:2019omf}. Here $ i =L, R $ correspond to the chiralities of the lepton current, and $\lambda=t,\pm1,0 $ represents the helicity state of the virtual gauge boson that decays into the dilepton pair. The symbols $\lambda_{1,2} $ are the helicities of the leptons. Additionally, the parameters $\eta_{t}$ and $\eta_{\pm1,0}$ are assigned a value of 1 and -1, respectively. The expressions of $H_{VA,\lambda}^{i,s_p,s_k}$ and  $H_{SP,\lambda}^{i,s_p,s_k}$ in terms of Wilson coefficients (WCs) and form factors (FFs) can be found in \cite{Das:2018sms}. Alternatively, in the literature, transversity amplitudes: $A_{\perp(||)1}^{i},~ A_{\perp(||)0}^{i}$ and $A_{S\perp(||)}, ~A_{P\perp(||)}$ are often employed instead of the hadronic helicity amplitudes. The expressions for these transversity amplitudes can be found in Ref. \cite{Das:2019omf}. The  amplitudes $L_{i,\lambda}^{\lambda_2,\lambda_1}$ and  $L_{i}^{\lambda_2,\lambda_1}$ are defined as follows:
\begin{align}
\begin{split}
& L^{\lambda_2,\lambda_1}_{L(R)} = \langle \bar{\ell}_2(\lambda_2)\ell_1(\lambda_1) | \bar{\ell}_2 (1\mp\gamma_5) \ell_1 | 0\rangle\, , \\
\label{eq:Ldef2}
& L^{\lambda_2,\lambda_1}_{L(R),\lambda} = \bar{\epsilon}^\mu(\lambda) \langle \bar{\ell}_2(\lambda_2) \ell_1(\lambda_1) | \bar{\ell}_2 \gamma_\mu (1\mp\gamma_5) \ell_1 | 0\rangle\, .
\end{split}
\end{align}
 Here, $\epsilon^{\mu}$ represents the polarization vector of the virtual gauge boson that decays into the dilepton pair. The detailed calculations of $L_{i,\lambda}^{\lambda_2,\lambda_1}$ and  $L_{i}^{\lambda_2,\lambda_1}$ can be found in Ref. \cite{Das:2019omf}. Based on these definitions, one can derive the differential branching ratio of $\Lambda_b \rightarrow \Lambda \ell_1\ell_2 $ as follows:
\begin{equation}
\begin{split}
\frac{d^{2}{\cal B}}{dq^2d\cos\theta_l}=\frac{3}{2}(K_{1ss}\sin^2\theta_l+K_{1cc}\cos^2\theta_l+K_{1c}\cos\theta_l)\;.
\end{split}
\end{equation}
The angle $\theta_{\ell}$ can vary within the range $-\pi \le \theta_{\ell}\le \pi$.
The long-distance aspect of the decay is encapsulated within the $ \Lambda_b \rightarrow \Lambda $ transition matrix elements, which are parameterized in terms of six $ q^2 $-dependent form factors, denoted as $f_{t,0,\perp}^{V/A}$ \cite{Feldmann:2011xf}. For our numerical analysis, we utilize the form factors obtained from lattice QCD calculations \cite{Detmold:2016pkz}. The resulting differential branching ratio is given by:
\begin{equation} \label{eq:diffBr}
	\frac{d\mathcal{B}}{dq^2} = 2 K_{1ss} + K_{1cc}\, .
\end{equation}
Additionally, the FBA (Forward-backward asymmetry) is expressed as,
\begin{equation}\label{eq:AlFB}
A^{\ell}_{\rm FB} = \frac{3}{2} \frac{K_{1c}}{K_{1ss} + K_{1cc}}\, ,
\end{equation}
where the squared  dilepton invariant mass ($q^2$)  varies within the range $(m_{1}+m_{2})^{2} \le q^{2} \le (m_{\Lambda_{b}} - m_{\Lambda})^2$.

\subsubsection{\textbf{Exclusive $B \to K_{2}^{*} \ell_{1}\ell_{2}$ decay channel}}
Utilizing the effective Hamiltonian governing $b \to s \ell_1 \ell_2$ transition, provided  in Eq. (\ref{eq:234}), one can derive the transition amplitude for the $B \to K_{2}^{*} \ell_{1}\ell_{2}$ decay mode. The hadronic matrix elements of vector and axial-vector currents for $B \to K_{2}^{*}$ transitions can be parameterized in terms of $q^2$-dependent FFs: $V(q^2)$ and  $A_{0,1,2}(q^2)$. The expressions in detail are given as follows \cite{Wang:2010ni},
 \begin{eqnarray}
  \langle K_2^*(k, \epsilon^*)|\bar s\gamma^{\mu}b|\overline B(p)\rangle
  &=&-\frac{2V(q^2)}{m_B+m_{K_2^*}}\epsilon^{\mu\nu\rho\sigma} \epsilon^*_{T\nu}  p_{\rho}k_{\sigma}, \nonumber\\
  \langle  K_2^*(k,\epsilon^*)|\bar s\gamma^{\mu}\gamma_5 b|\overline B(p)\rangle
   &=&i(m_B+m_{K_2^*})A_1(q^2)\left[ \epsilon^{*\mu}_{T}
    -\frac{\epsilon^*_{T } \cdot  q }{q^2}q^{\mu} \right]+2im_{K_2^*} A_0(q^2)\frac{\epsilon^*_{T } \cdot  q }{ q^2}q^{\mu}  \nonumber\\
    &&-iA_2(q^2)\frac{\epsilon^*_{T} \cdot  q }{  m_B+m_{K_2^*} }
     \left[ (p+k)^{\mu}-\frac{m_B^2-m_{K_2^*}^2}{q^2}q^{\mu} \right],
\end{eqnarray}
where, $p$ ($k$) is the four momentum of $B$ ($K_{2}^{*}$)  meson.  

We employ the recent values of form factors from the light cone QCD sum rule (LCSR) approach given in Ref. \cite{Aliev:2019ojc}.
In this method, the form factors can be expressed as 
\begin{equation}
    F^{B \to T}=\frac{1}{1 - q^{2}/ m_{R,F}^{2}} \sum_{n=0}^{1}\alpha_{n}^{F}[z(q^{2})-z(0)]^{n},
\end{equation}
where $z(q^{2})=\frac{\sqrt{t_{+}-q^{2}}- \sqrt{t_{+}-t_{0}}}{\sqrt{t_{+}-q^{2}}+ \sqrt{t_{+}-t_{0}}}$, ~~$t_{\pm}= (m_{B}{\pm} m_{K_{2}^{*}})^2$, ~~$t_{0}=t_{+}(1-\sqrt{1-t_{-}/t_{+}})$ and $m_{R,F}$ is the resonance mass corresponding to the form factor. The associated parameters are provided in Table \ref{tab:K2para}. The resonance masses employed in our numerical calculations are given as
\begin{eqnarray}
m_{R,A_{0}}=5.336 \hspace{0.1cm}{\rm GeV}, \hspace{0.5cm} m_{R,V}=5.412 \hspace{0.1cm}{\rm GeV}, \hspace{0.5cm} m_{R,(A_1,A_3)}=5.829 \hspace{0.1cm}{\rm GeV}.
\end{eqnarray}

\begin{table}[tbp]
\centering
\begin{tabular}{|c|c|c|}
\hline
Form factor &  $\alpha_{0}$ &$\alpha_{1} $\\
\hline
$ V^{B\to K_{2}^{*}}$ & $0.22^{+0.11}_{-0.08}$ & $-0.90^{+0.37}_{-0.50}$  \\
$ A^{B\to K_{2}^{*}}_{0}$ & $0.30^{+0.01}_{-0.05}$ & $-1.23^{+0.23}_{-0.23}$  \\
$ A^{B\to K_{2}^{*}}_{1}$ & $0.19^{+0.09}_{-0.07}$ & $-0.46^{+0.19}_{-0.25}$  \\
$ A^{B\to K_{2}^{*}}_{2}$ & $0.11^{+0.05}_{-0.06}$ & $-0.40^{+0.23}_{-0.16}$  \\
\hline 
\end{tabular}
\caption{ Fit parameters corresponding to $B \to K_{2}^{*}$ form factors in LCSR method }
\label{tab:K2para}
\end{table}

The differential decay distribution describing  three body $B \to K_2^*\ell_1 \ell_2$ decay can be expressed as \cite{Kumbhakar:2022szr} 
\begin{equation}\label{eq:2222}
    \frac{d^{2} \Gamma}{dq^{2} d\cos\theta_{\ell}}=A(q^{2})+B(q^2)\cos\theta_{\ell}+C(q^2)\cos^{2}\theta_{\ell}\;,
\end{equation}
where $\theta_{\ell}$ is the leptonic polar angle which describes the angle made by lepton $\ell_{1}$ to the dilepton rest frame. The $q^2$-dependent coefficients $A(q^{2}), B(q^2)$, and $C(q^2)$ are given in Appendix -\ref{inputsBtoK2star}.
Using Eq.~(\ref{eq:2222}), the differential decay rate can be given as
\begin{equation}
    \frac{d\Gamma}{d q^2}= 2\left(A + \frac{C}{3}\right),
\end{equation}
and the lepton FBA is found to be
\begin{equation}
A_{\rm FB}(q^2)= \frac{1}{d\Gamma/dq^2}\left(\int_0^1 d\cos\theta_\ell\frac{d^2\Gamma}{d\cos\theta_\ell d q^2}-\int_{-1}^0 d\cos\theta_\ell \frac{d^2\Gamma}{d\cos\theta_\ell d q^2 }\right) = \frac{B}{2\left(A+\frac{C}{3}\right)}\;.
\end{equation}

\subsubsection{\textbf{Exclusive $B \rightarrow (K^*,\phi) \ell_1^-\ell_2^+$ decay channel}}
\label{subsec:kphi}
For the analysis of the decay $B \rightarrow (K^*,\phi) \ell_1^-\ell_2^+$, we adopt the kinematics and the angular conventions as described in references \cite{Korner:1989qb, Becirevic:2016zri}.
The hadronic matrix elements involve a more extensive set of $q^2$-dependent form factors which include
\begin{align}\label{def:FFV}
\langle \bar{K}^\ast(k)|\bar{s}\sigma_{\mu\nu} q^\nu(1-\gamma_5) b|\bar{B}(p)\rangle &= 2 i \varepsilon_{\mu\nu\rho\sigma} \varepsilon^{\ast\nu}p^\rho k^\sigma T_1(q^2)-[(\varepsilon^\ast \cdot q)(2p-q)_\mu - \varepsilon_\mu^\ast(m_B^2-m_{K^\ast}^2)]T_2(q^2)\nonumber\\[.3em] 
&-(\varepsilon^\ast \cdot q)\Big{[} \frac{q^2}{m_B^2-m_{K^\ast}^2}(p+k)_\mu - q_\mu \Big{]}T_3(q^2), \nonumber\\[.7em] 
\langle \bar{K}^\ast(k)|\bar{s}\gamma^\mu(1-\gamma_5) b|\bar{B}(p)\rangle &= \varepsilon_{\mu\nu\rho\sigma}\varepsilon^{\ast\nu}p^\rho k^\sigma \frac{2 V(q^2)}{m_B+m_{K^\ast}}-i\Big{[} \varepsilon_\mu^\ast(m_B+m_{K^\ast})A_1(q^2)\nonumber\\[.3em] 
&-i(p+k)_\mu (\varepsilon^\ast \cdot q)\frac{A_2(q^2)}{m_B+m_{K^\ast}}- q_\mu(\varepsilon^\ast \cdot q) \frac{2 m_{K^\ast}}{q^2}[A_3(q^2)-A_0(q^2)]\Big{]}.
\end{align}
The polarisation vector of the $K^{*}(\phi)$ meson is denoted as $\varepsilon_{\mu }$. The form factor $A_{3}(q^2)$ related to $A_1(q^2)$ and $A_2(q^2)$ is given by $2m_{V}A_{3}(q^2)=(m_{B}+m_{V})A_{1}(q^2)-(m_{B}-m_{V})A_{2}(q^2)$.

The $q^2$-dependent  diffrential branching ratio, after integrating the full angular distribution over the angles given in Appendix \ref{InputsBtoV}, are expressed as
\begin{equation}
\begin{split}
\frac{d{\cal B}}{dq^2}=\frac{1}{4}[3I_1^c(q^2)+6I_1^s(q^2)-I_2^c(q^2)-2I_2^s(q^2)]
\end{split}
\end{equation}
Similarly, the forward-backward asymmetry and lepton polarisation asymmetry, are given as follows
\begin{equation}
A_{\rm FB}(q^2)=\frac{3I_6^s (q^2)+3/2 I_6^c (q^2)}{3I_1^c(q^2)+6I_1^s(q^2)-I_2^c(q^2)-2I_2^s(q^2)},
\end{equation}
\begin{equation}
F_{\rm L}(q^2)=\frac{3I_1^c (q^2)- I_2^c (q^2)}{3I_1^c(q^2)+6I_1^s(q^2)-I_2^c(q^2)-2I_2^s(q^2)}.
\end{equation}
 The form factors associated with the transversity amplitude are obtained using the LCSR method  \cite{Bharucha:2015bzk} and are expressed as follows:
\begin{equation}
    F_{i}(q^{2})=\frac{1}{1 - q^{2}/ m_{R,i}^{2}} \sum_{k}\alpha_{k}^{i}[z(q^{2})-z(0)]^{k},
\end{equation}
where $z(q^{2})=\frac{\sqrt{t_{+}-q^{2}}- \sqrt{t_{+}-t_{0}}}{\sqrt{t_{+}-q^{2}}+ \sqrt{t_{+}-t_{0}}}$,~ $t_{\pm}= (m_{B}{\pm} m_{V})^2$, ~$t_{0}=t_{+}(1-\sqrt{1-t_{-}/t_{+}})$ and $m_{R,F}$ is the resonance mass corresponding to the form factor. In our calculations, we use the resonance masses as follows: 
\begin{eqnarray}
m_{R,A_{0}}=5.336 \hspace{0.1cm}{\rm GeV}, \hspace{0.5cm} m_{R,(V,T_1)}=5.412 \hspace{0.1cm}{\rm GeV}, \hspace{0.5cm} m_{R,(A_1,A_3,T_2,T_3)}=5.829 \hspace{0.1cm}{\rm GeV}.
\end{eqnarray}.
 We apply the above formalism originally developed for $B \to K^{*} $ decay to analyze the process $B \to \phi \ell_1 \ell_2$. This is achieved by straightforwardly substituting the relevant mass and form factors for the corresponding vector meson $\phi$. The numerical values for the parameters associated with the form factors with $1\sigma$ uncertainty are given in Table \ref{tab:Formfactors for k}.
\begin{table}[tbp]
\centering
\begin{tabular}{|c|c|c|c|}
\hline

$F_{i}$ & $\alpha_{0}$ & $\alpha_{1}$ &$ \alpha_{2}$ \\
\hline
$V$ & $0.34 \pm 0.04$ & $-1.05\pm 0.24$& $2.37\pm 1.39$\\
$ A_{0}$ &$0.36\pm 0.05$ &$-1.04\pm 0.27$&$1.12\pm1.35$ \\
$A_1$ &$0.27\pm0.03$ & $0.39\pm0.19$ &$-0.11\pm0.48$ \\
$ A_3$ &$0.26\pm0.03$ & $0.60\pm0.20$ &$0.12\pm0.84$ \\
$ T_1$ & $0.28\pm0.03$ &$ -0.89\pm0.19$ &$1.95\pm1.10$ \\
$ T_2$ &$0.28\pm0.03$& $0.40\pm0.18$ &$0.36\pm0.51$ \\
$T_3$ &$0.67\pm0.08$ & $1.48\pm0.49$ &$1.92\pm1.96$ \\
\hline 
\end{tabular}
\hspace{7mm}
\begin{tabular}{|c|c|c|c|}
\hline
$F_{i}$ & $\alpha_{0}$ & $\alpha_{1}$ &$ \alpha_{2}$ \\
\hline
$V$ & $0.39 \pm 0.03$ & $-1.03\pm0.25$& $3.50\pm1.55$\\
$ A_{0}$ &$0.39\pm0.05$ &$-0.78\pm0.26$&$2.41\pm1.48$ \\
$A_1$ &$0.30\pm0.03$ & $0.48\pm0.19$ &$0.29\pm0.65$ \\
$ A_3$ &$0.25\pm0.03$ & $0.76\pm0.20$ &$0.71\pm0.96$ \\
$ T_1$ & $0.31\pm0.03$ &$ -0.87\pm0.19$ &$2.75\pm1.19$ \\
$ T_2$ &$0.31\pm0.03$& $0.58\pm0.19$ &$0.89\pm0.71$ \\
$T_3$ &$0.68\pm0.07$ & $2.11\pm0.46$ &$4.94\pm2.25$ \\
\hline 
\end{tabular}
\caption{Form factors for $B \to K^{*}\ell_1 \ell_2$ (left panel) and $B_{s} \to \phi \ell_1 \ell_2$ (right panel).}
\label{tab:Formfactors for k}
\end{table} 
\section{phenomenological implication}
\label{subsec:pheno}
SMEFT-inspired   LFV transitions   have garnered significant attention. Given the absence of any new particles observed thus far beyond the electroweak scale, it is conjectured that the scale of NP is substantially higher compared to the current running scale of the LHC. In this context, the SMEFT offers a powerful framework for elucidating LFV decays which encompasses a comprehensive set of dimension-six operators constructed from the fields of the Standard model. The corresponding Lagrangian is given as follows \cite{Grzadkowski:2010es}:
\begin{equation}
    \begin{split}
        \mathcal L_{SMEFT}&=\mathcal L_{SM}-\frac{1}{\Lambda_{\rm cut}^2}\Big\{[C^{(3)}_{lq}]^{ij\alpha \beta}(\bar{Q}^i \gamma^{\mu}\sigma^{a}Q^{j})(\bar{L}^\alpha\gamma_{\mu}\sigma^{a}L^{\beta})\\&+[C^{(1)}_{lq}]^{ij\alpha \beta}(\bar{Q}^i \gamma^{\mu}Q^{j})(\bar{L}^\alpha\gamma_{\mu}L^{\beta})+[C_{leqd}^{ij\alpha\beta}(\bar{Q}^id_{R}^j)(\bar{e}_{R}^\alpha L^{\beta})]\Big \} +{\rm  h.c}.
    \end{split}
\end{equation}
Here $Q$ and $L$ represent the  left-handed quark and lepton fields,  which transform as doublets under $SU(2)$ whereas $e_{R}$ and $d_{R}$ are the singlet right-handed charged leptons and down-type quarks, respectively. 
Here   $ \Lambda_{\rm cut} $ is the cut-off scale which can be associated with the mass of the heavy NP degrees of freedom. 
The above equation includes the set of all the dimension six operators contributing to the $b \to s\ell_{1}\ell_{2}$ transitions. It should be noted that none of these above operators contain tensor currents.

The SMEFT Wilson coefficients can be constrained from low energy processes, and are related to the Wilson coefficients in Eqn (\ref{eq:234}) as
\bea
        &&C_9^{\ell_1\ell_2}=-C_{10}^{\ell_1\ell_2}=+\frac{v^2}{\Lambda_{\rm cut}^2}\frac{\pi}{\alpha_{em}|{V_{tb}V^{*}_{ts}}|}([C^{(3)}_{\ell q}]^{23\ell_1\ell_2}+[C^{(1)}_{\ell q}]^{23\ell_1\ell_2}),\\
      &&  C_9^{'\ell_1\ell_2}=-{C'}_{10}^{\ell_1\ell_2}=+\frac{v^2}{\Lambda_{\rm cut}^2}\frac{\pi}{\alpha_{em}{V_{tb}V^*_{ts}}|}([C_{\ell d}]^{23\ell_1\ell_2},\\ &&C_S^{\ell_1\ell_2}=-C_{P}^{\ell_1\ell_2}=+\frac{v^2}{\Lambda_{\rm cut}^2}\frac{\pi}{\alpha_{em}{V_{tb}V^*_{ts}}|}([C_{leqd}]^{23\ell_1\ell_2},\\ 
      && C_S^{'\ell_1\ell_2}=-{C'}_{P}^{\ell_1\ell_2}=+\frac{v^2}{\Lambda_{\rm cut}^2}\frac{\pi}{\alpha_{em}{V_{tb}V^*_{ts}}|}([C_{\ell eqd}^*]^{32\ell_1\ell_2} \;.
\eea
Now, we examine the constraints on various combinations of SMEFT Wilson coefficients derived from measurements of mesonic LFV decays. We consider the branching ratios of the decay modes $\bar{B}_{s}\rightarrow \ell_{1}^{-}\ell_{2}^{+}$ and $B\rightarrow K \ell_{1}^{-}\ell_{2}^{+}$ and use  their experimental upper limits provided in Table \ref{tab:LQs} at 90$\%$ confidence level.
\begin{table}[htb]
\centering
\begin{tabular}{|c|c|c|c|}
\hline
Observable  & Exp.limit\\
\hline
$ \mathcal{B}(B_{s} \to \mu^{\pm} \tau^{\mp})$  & $ 4.2 \times 10^{-5}$ \cite{LHCb:2019ujz} \\
$\mathcal{B}(B^{+} \to K^{+} \mu^{-} \tau^{+})$ & $3.9 \times 10^{-5}$ \cite{LHCb:2020khb}\\
$\mathcal{B}(B^{+} \to K^{+} \mu^{+} \tau^{-})$ & $4.5 \times 10^{-5}$ \cite{BaBar:2012azg}\\
\hline 
\end{tabular}
\caption{Experimental upper limits for LFV $B$ decays at 90$\%$ C.L.} 
\label{tab:LQs}
\end{table} 

The branching fraction of these decay modes are expressed as \cite{Becirevic:2016oho, Gratrex:2015hna}
\begin{equation}\label{eq:BrBs2ll}
\begin{aligned}
\mathcal{B}(\bar{B}_s\to\ell^-_1 \ell^+_2)=&\,\frac{\tau_{B_s}}{64\pi^3}\frac{\alpha_\text{em}^2G_F^2 |V_{tb}V_{ts}^*|^2}{m_{B_s}^3} f_{B_s}^2 \, \lambda^{1/2}(m_{B_s}^2,m_{\ell_1}^2,m_{\ell_2}^2) \\
\times&\left\{[m_{B_s}^2-(m_{\ell_1}-m_{\ell_2})^2]\left|(m_{\ell_1}+m_{\ell_2}){C}_{10-}+\frac{m_{B_s}^2}{m_b+m_s}{C}_{P-}\right|^2\right. \\
&+\left.[m_{B_s}^2-(m_{\ell_1}+m_{\ell_2})^2]\left|(m_{\ell_1}-m_{\ell_2})({C}_{9-})+\frac{m_{B_s}^2}{m_b+m_s}({C}_{S-})\right|^2\right\} ,
\end{aligned}
\end{equation}

 \begin{equation}\label{eq:BrB2Kll}
\begin{aligned}
\mathcal{B}(B^+\to K^+\ell_1^-\ell_2^+) &= 10^{-8} \bigg\{ c^S_{\ell_1\ell_2}\left|C_{S+}\right|^2+c^P_{\ell_1\ell_2}\left|C_{P+}\right|^2+c^{9+}_{\ell_1\ell_2}\left|C_{9+} \right|^2  \\[2pt]
 & + c^{10+}_{\ell_1\ell_2}\left|C_{10+} \right|^2   +c^{S9}_{\ell_1\ell_2}\,\mathrm{Re}[C_{S+}^{*}  C_{9+}] + c^{P10}_{\ell_1\ell_2}\,\mathrm{Re}[C_{P+}^{*} C_{10+}]
 \bigg\}\,, 
\end{aligned}
\end{equation}
where we have adopted the notation $C_{X\pm}=C_{X}\pm C_{X}^{\prime}$.
To calculate the branching ratios for these processes,  we use the values of particle masses from PDG \cite{ParticleDataGroup:2020ssz}, CKM matrix elements from the UT-fit collaborations \cite{utfit}, and the decay constant of $B_s$ meson as $f_{B_{s}}$=215 MeV \cite{Balasubramamian:2019wgx}.
The values of the coefficients $c^{i}_{\ell_1\ell_2}$ are taken from \cite{Bordone:2021usz}.
For our analysis, we utilize the Eqs. (\ref{eq:BrBs2ll}) and (\ref{eq:BrB2Kll}) to constrain the various combinations of SMEFT Wilson coefficients. For convenience, we sum over the oppositely charge lepton decays modes e.g., $\mathcal{B}(B_{s} \to \mu^{\pm} \tau^{\mp})=(B_{s}\to \mu^{-}\tau^{+})+(B_{s}\to \mu^{+}\tau^{-})$ and similarly for other channel as well. Here we consider the constraints coming from the $\mu \tau $ decay channel. It is generally established that consideration of primed operators is unappealing to fit the $b \to s \ell \ell $ data \cite{Alguero:2021anc, Altmannshofer:2021qrr, Hurth:2021nsi, Ciuchini:2019usw, Gratrex:2015hna} and hence, the use of only unprimed operators is well accepted. 
In our analysis, we set the cutoff scale to be 1 TeV. Now we perform the $\chi^{2}$ analysis to obtain the NP parameter space allowed by the current data.
The  $\chi ^2$ function  is defined as
\bea
\chi^2(C^{\rm NP})= \sum_i  \frac{\Big ({\cal O}_i^{\rm Th}(C^{\rm NP}) -{\cal O}_i^{\rm Exp} \Big )^2}{(\Delta {\cal O}_i^{\rm Exp})^2+(\Delta {\cal O}_i^{\rm SM})^2},
\eea
where ${\cal O}_i ^ {\rm Th}$ and ${\cal O}_i ^ {\rm Exp}$ represent the theoretical values and the measured central value of the observables, respectively. The denominator represents the error associated with the Standard Model and experimental values.
The measured value and the upper limits of each observable are listed in Table \ref{tab:LQs} which are incorporated into the fit. We note that there are only a 90$\%$ C.L for the upper limits on the branching ratios of $B^{+} \to K^{+}\mu^{\pm}\tau^{\mp}$ and $B_{s} \to \mu^{\pm}\tau^{\mp}$. In order to incorporate these observables into our fit, we take the branching ratio to be (0.0$\pm$U.L/1.645). These decays are highly suppressed in the Standard Model so the error associated with the SM value is considered to be zero. The allowed NP couplings are obtained by minimizing the $\chi^{2}$ function. We obtain the allowed new physics parameter space in the planes of $[C_{leqd}]^{23\tau \mu}$ - $[C^{(1)}_{\ell q}]^{23\tau \mu}$,  where the red, green and blue regions represent the $1\sigma$,  $2\sigma$, and $3\sigma$ contours around the $\chi^{2}_{\rm min}$ value. The regions are shown in Fig \ref{fig::ModelIndep-constrainstau}, where
we illustrate the constrained parameter space governing the SMEFT Wilson coefficients for the final states involving $\tau \mu$. Here we focus on the two-dimensional scenario involving the SMEFT new physics couplings $ [C^{(1)}_{\ell q}]^{23\tau \mu}$ and $[C_{leqd}]^{23\tau \mu}$ which enables us to elucidate the intricate relationship between these coefficients within the SMEFT framework.

\section{\textbf{Implications  OF the Results}}
\label{subsec:result}
\begin{figure}
\begin{center}
\hspace{2mm}
\includegraphics[height=65 mm,width=75mm]{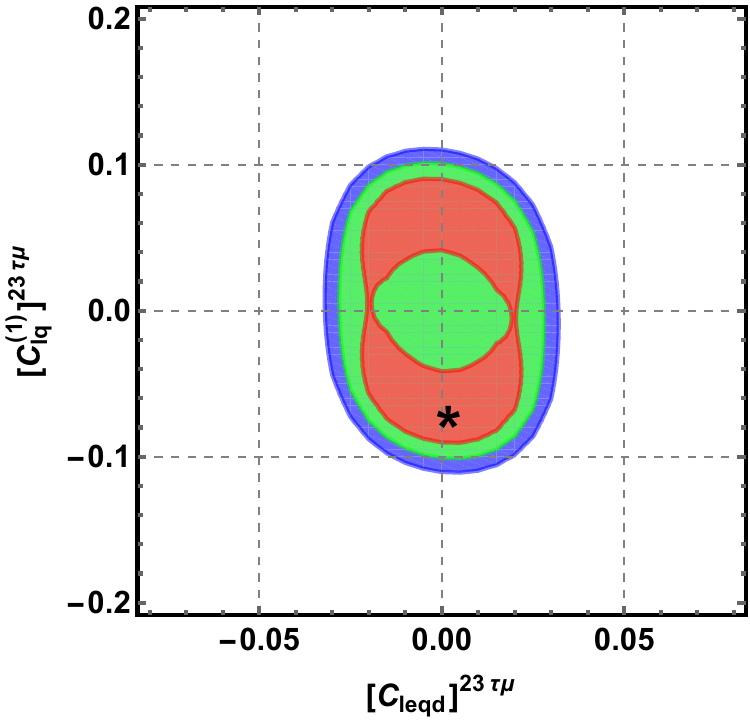}
\end{center}
\vspace{-6mm}
\small \caption{Constraints on the NP couplings obtained from the combined measurement of ${\cal B}(B \to \tau \mu$) and ${\cal B}(B \to K^{+} \tau \mu$), where the red, green and blue regions represent the $1\sigma$,  $2\sigma$, and $3\sigma$ contours, while the black star indicating the best-fit value.}
 \label{fig::ModelIndep-constrainstau}
\end{figure}   
In this section, we illustrate the implications of our results in  $b \to s \ell _1 \ell _2$ transitions. The allowed parameter space is constrained by using the combined measurements of  $B_{s} \to \tau^{\pm} \mu^{\mp}  $ and $B \to K^{+}\tau^{\pm} \mu^{\mp}$,  experimental upper limits set at 90$\%$  C.L. is shown in Fig. \ref{fig::ModelIndep-constrainstau}. Here we focus only on the $\tau \mu $ mode. 
For our computation, we set the cut-off scale  $\Lambda_{\rm cut} =1 $ TeV and use the best-fit values of the Wilson coefficients as $ [C^{(1)}_{\ell q}]^{23\tau \mu}=-0.0705$  and $[C_{leqd}]^{23\tau \mu}=0.0019$.

Using these values, we analyze the $q^2$ dependencies of the key observables such as the branching fraction, the lepton forward-backward asymmetry, and the longitudinal polarization fractions in the discussed $b \to s \ell _1 \ell _2$ transitions. The plots presented in Sec. \ref{subsec:result} utilize the 1$\sigma$ standard error values of the form factors.
\subsection{Impact of SMEFT NP coefficients on $\Lambda_{b} \to \Lambda \ell_1 \ell_2$  decay observables}
\begin{figure}[h!]
\begin{center}
\includegraphics[height=55mm,width=75mm]{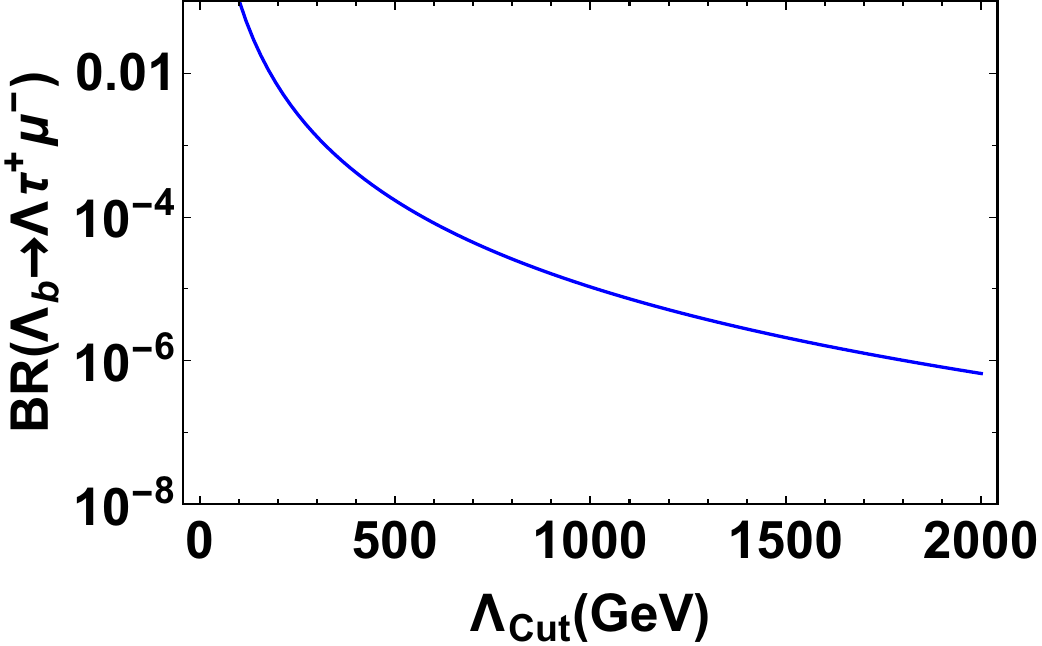}
\hspace{2mm}
\includegraphics[height=55mm,width=75mm]{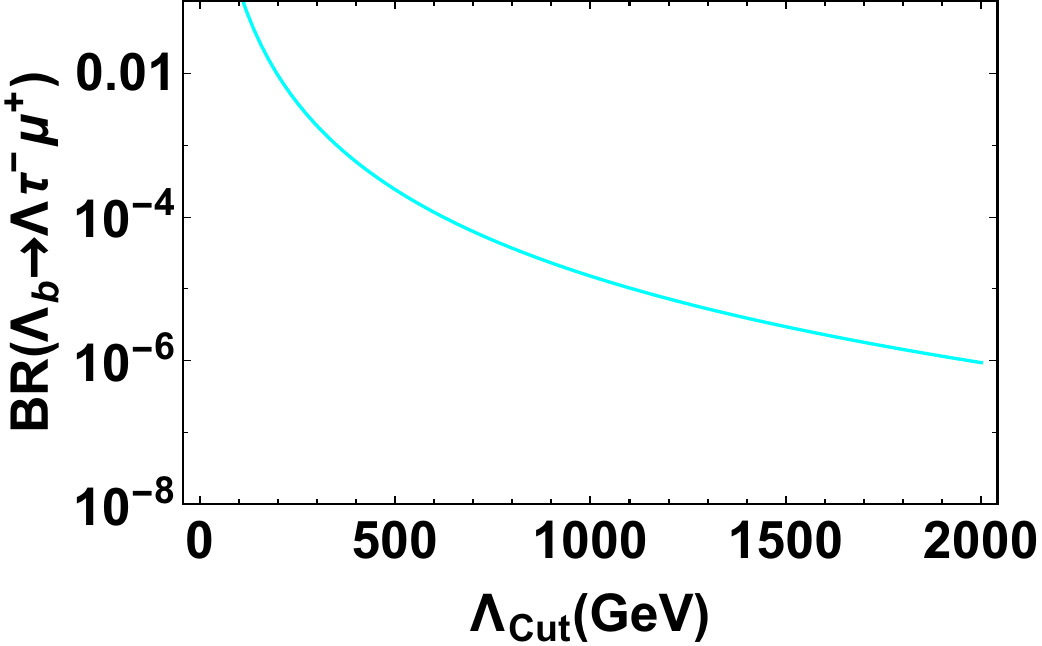}
\end{center}
\caption{ Variation of the branching ratio of $\Lambda_b\to\Lambda \tau^{+} \mu^{-} $ (left), $\Lambda_b\to\Lambda\tau^{-} \mu^{+} $ (right) with the cut-off scale. }
\label{fig:cutoffbrL}
\end{figure}

\begin{itemize}
   \item  $\textbf{Branching ratio}$: Considering the $\Lambda_b \to \Lambda \tau^{\pm} \mu^{\mp}$ decay process, the $q^2$-dependent differential branching ratio is depicted in the Fig \ref{fig:BRLL}. In the high $q^2$ region the decay rate distribution indicates that the contribution of the new Wilson coefficient is quite substantial. The solid black line in these plots represents the value of the differential branching ratio considering the central values of the form factors. The band on both sides of the central solid line indicates $1\sigma $ uncertainty for the corresponding observable.
\begin{figure}[htb]
\centering
\includegraphics[height=53mm,width=73mm]{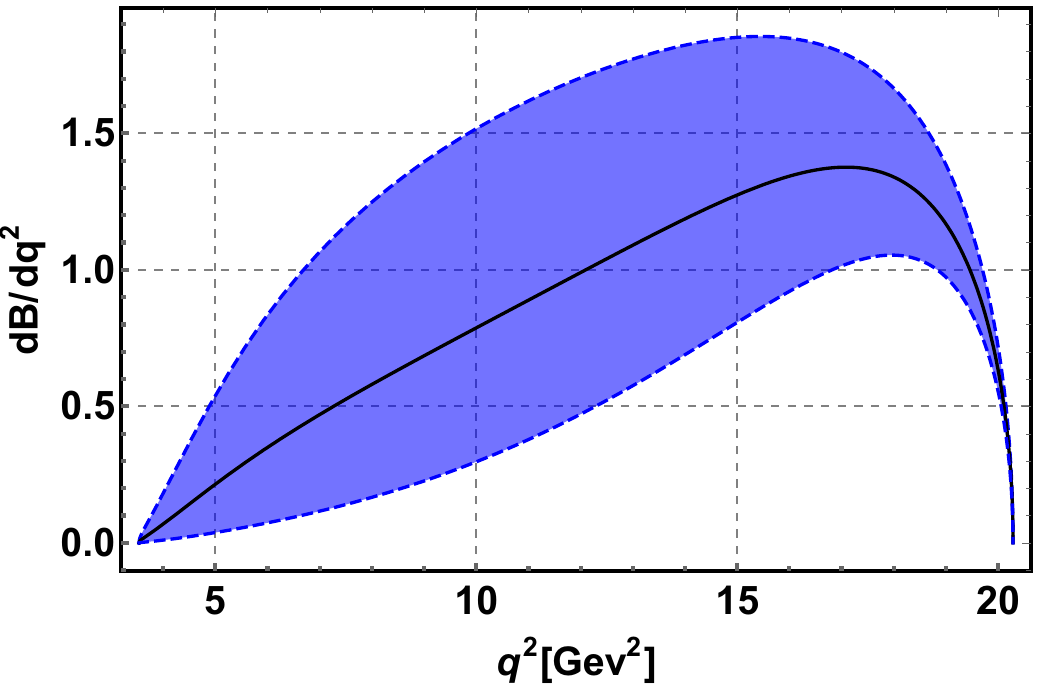}
\hspace{2mm}
\includegraphics[height=53mm,width=73mm]{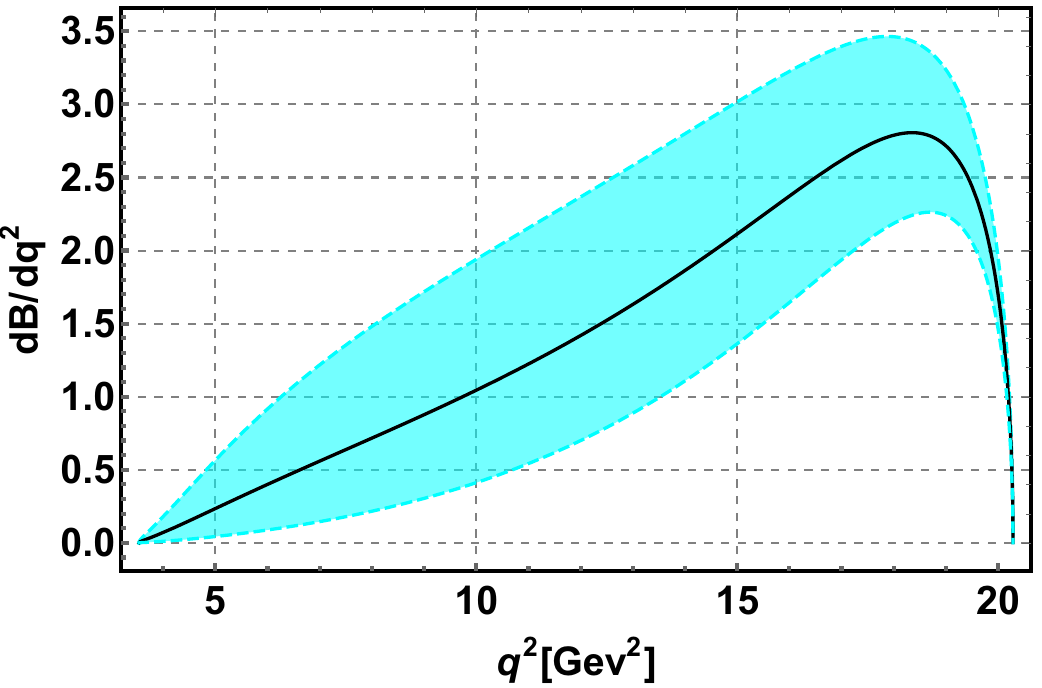}
\caption{Branching ratio (in units of $10^{-6}$) of  $\Lambda_{b} \to  \Lambda \tau^{+}\mu^{-}$ (left) and  $\Lambda_{b} \to  \Lambda \tau^{-}\mu^{+}$ (right).}
\label{fig:BRLL}
\end{figure}
   From the differential branching ratio plot, it is evident that both decay modes are expected to have similar order of branching fractions. However, the predicted value of $\mathcal{B}(\Lambda_{b}\to \Lambda \mu^{+}\tau^{-})$ is found to be slightly higher than that of $\mathcal{B}(\Lambda_{b}\to \Lambda \mu^{-}\tau^{+})$.
\end{itemize}
\vspace{-1mm}
\begin{itemize}
    \item  $\textbf{Forward-backward asymmetry}$: Taking into account the lepton forward-backward asymmetry, the $\Lambda_{b} \to \Lambda  \tau^{+} \mu^{-}$ decay exhibits a zero crossing in the FBA curve, whereas in the case of $\Lambda_{b} \to \Lambda  \tau^{-} \mu^{+}$,  no such zero crossing is observed. Such asymmetry curves are shown in Fig. \ref{fig::AFBLL}. The zero crossing for the former $\Lambda_{b} $ decay occurs at $q^2$=9 ${\rm GeV}^2$. For $\Lambda_{b} \to \Lambda  \tau^{-} \mu^{+}$,  the forward backward asymmetry remains negative throughout all $q^{2}$ regions. 
\begin{figure}[h!]
\centering
\includegraphics[height=53mm,width=73mm]{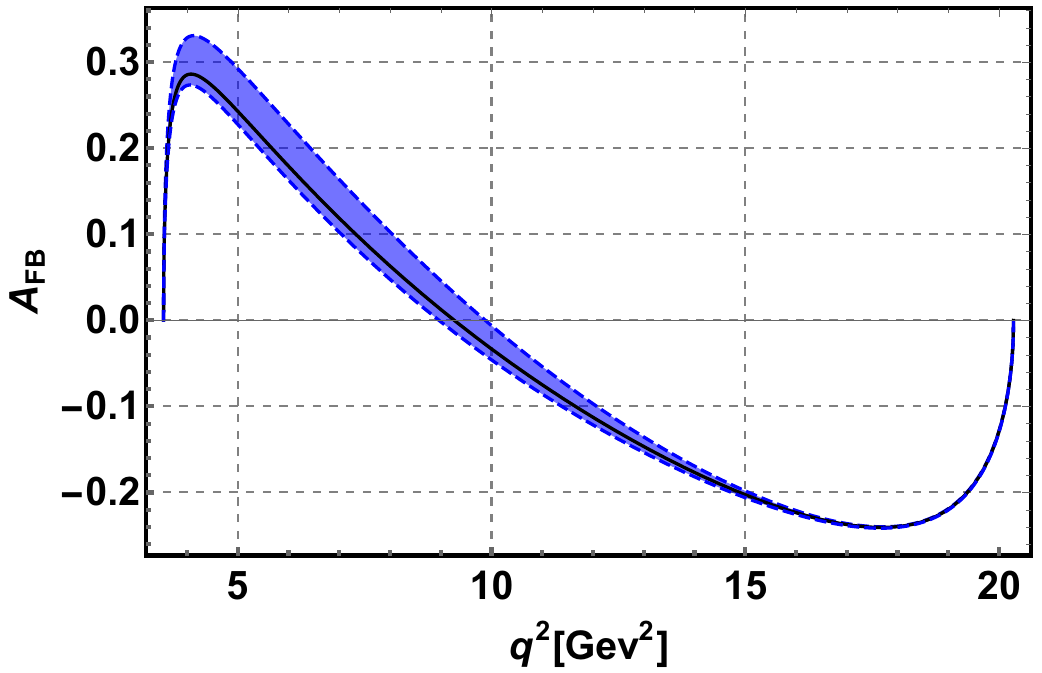}
\includegraphics[height=53mm,width=73mm]{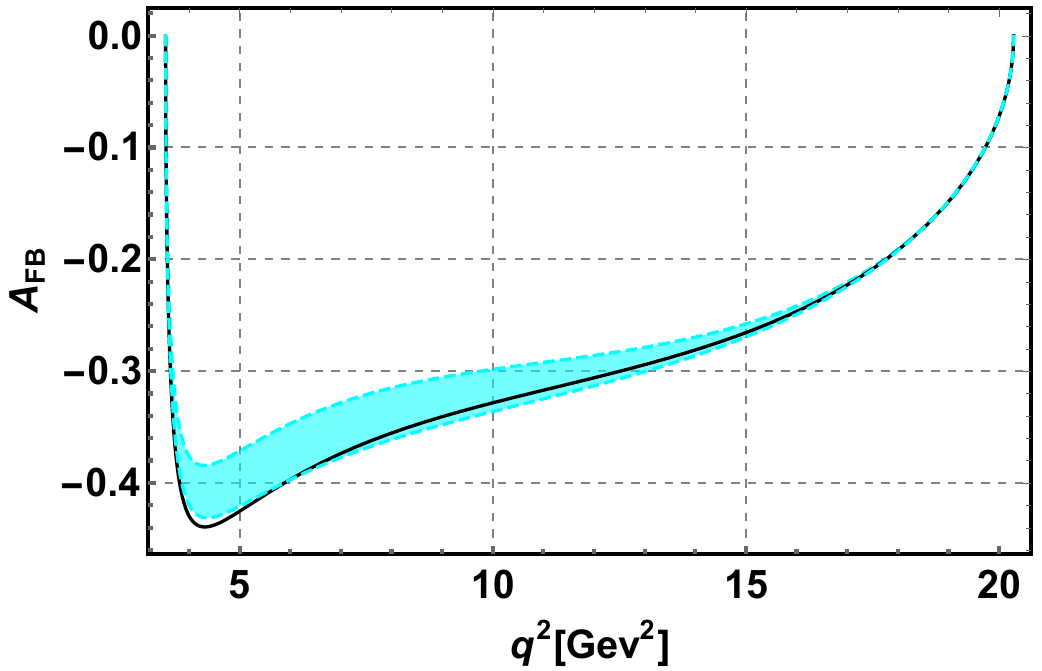}
\caption{Forward-backward asymmetry of  $\Lambda_{b} \to  \Lambda \tau^{+}\mu^{-}$ (left) and  $\Lambda_{b} \to  \Lambda \tau^{-}\mu^{+}$ (right).}
\label{fig::AFBLL}
\end{figure}
\end{itemize}
The numerical estimation for the branching ratio and $A_{FB}$ are computed using the central value of the form factors and best-fit points of the SMEFT Wilson coefficient. This has been shown in Table \ref{tab::Results}. As mentioned before  the obtained results correspond to the cut-off scale $\Lambda_{cut}$=1 TeV, and the evolution of the branching ratios of the $\Lambda_{b} \to \Lambda \mu^\pm \tau^\mp$ with the cut-off scale are shown in the Fig. \ref{fig:cutoffbrL}.
\subsection{Impact of SMEFT NP coefficients on $B \to K_{2}^{*} \ell_{1} \ell_{2}$  decay observables}
\begin{figure}[h!]
\begin{center}
\hspace{-6mm}
\includegraphics[height=55mm,width=75mm]{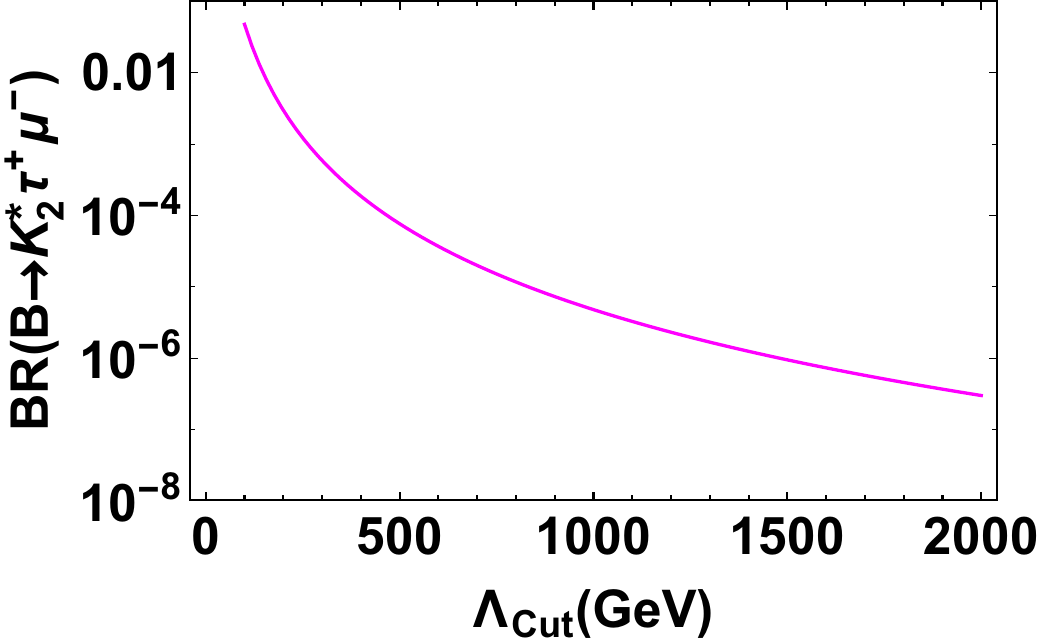}
\includegraphics[height=55mm,width=75mm]{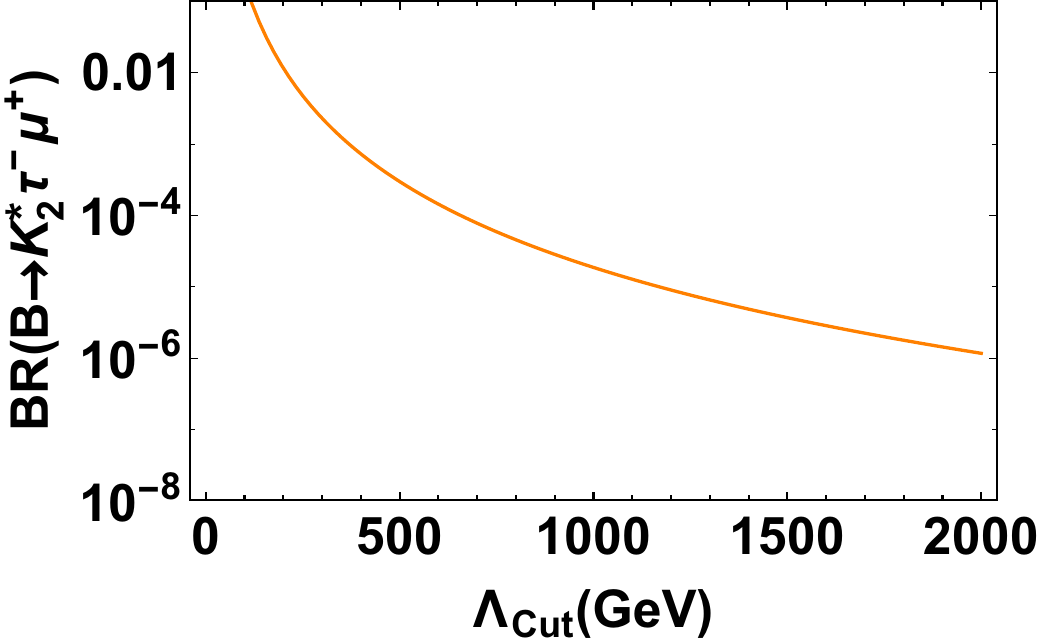}
\caption{Cut off dependent branching ratio of $B \to K_{2}^{*} \tau^{+} \mu^{-}$ (left) and $B \to K_{2}^{*} \tau^{-} \mu^{+}$ (right).}
\label{fig:cutoffbr11}
\end{center}
\end{figure}

\begin{itemize}
\item $\textbf{Branching ratio}$: Fig. \ref{fig:BRL} shows $q^{2}$-dependent differential branching ratios of the $B \to K_{2}^{*}\mu^{\pm}\tau^{\mp}$ processes. The contribution of SMEFT NP couplings is significant in the intermediate invariant mass squared region. The behavior of the differential branching ratios for both decay modes are of similar. However, the  value of  ${\cal B}(B \to K_{2}^{*}\mu^{-}\tau^{+})$ is higher than that for the  ${\cal B}(B \to K_{2}^{*}\mu^{+}\tau^{-})$.
    \begin{figure}[h!]
\centering
\includegraphics[height=53mm,width=73mm]{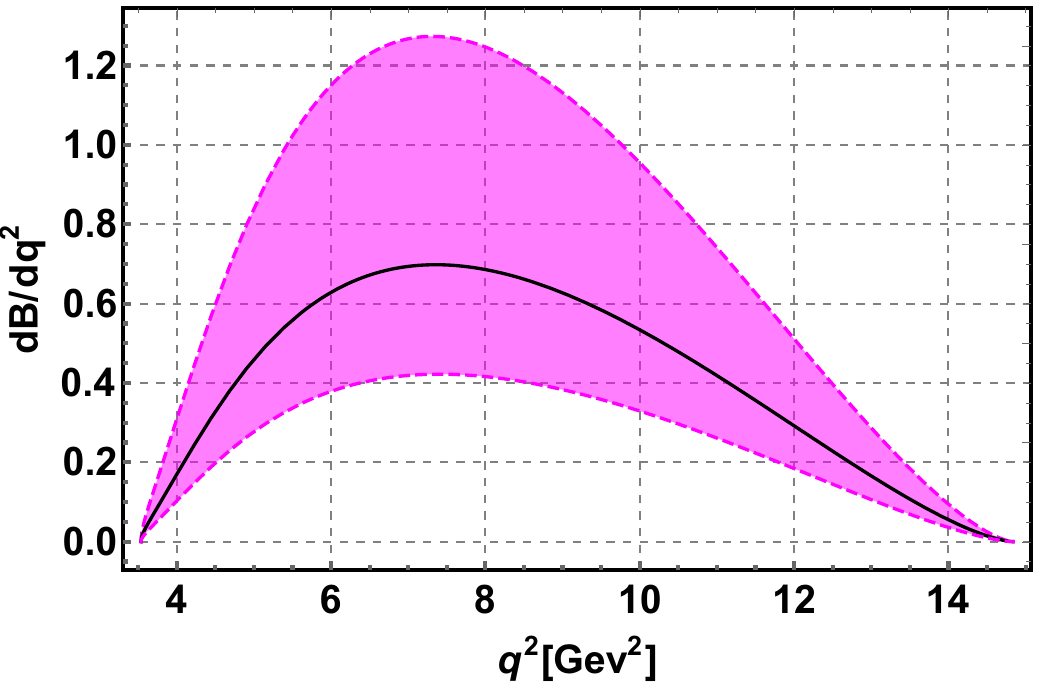}
\hspace{2mm}
\includegraphics[height=53mm,width=73mm]{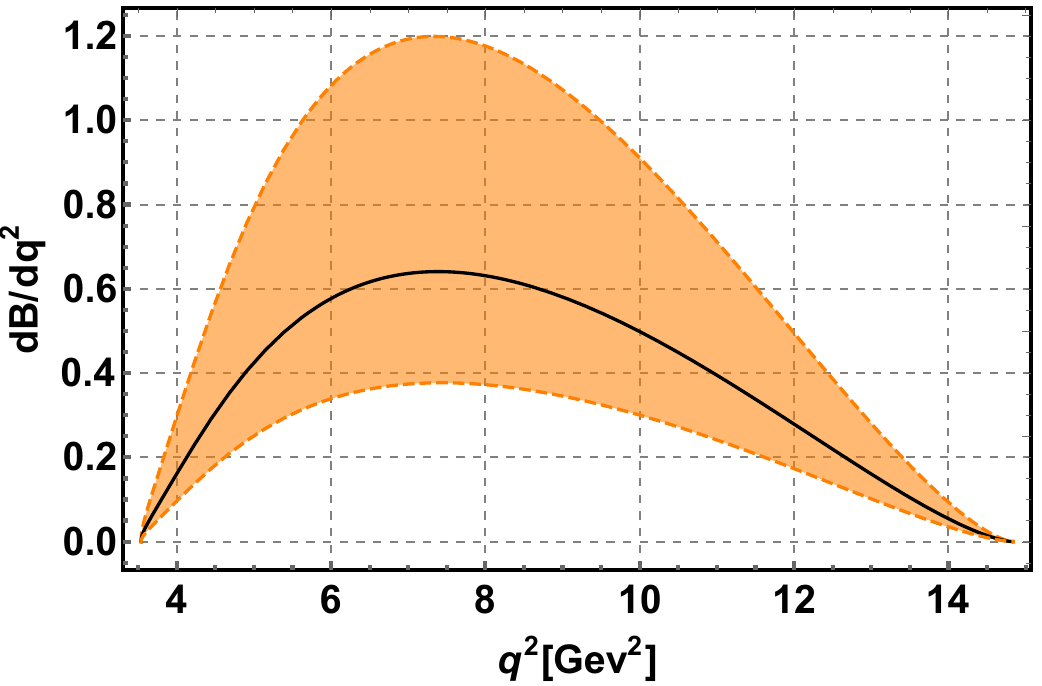}
\caption{Branching ratio (in units of $10^{-6}$) of  $B \to K_{2}^{*}\mu^{-}\tau^{+}$ (left) and $B \to K_{2}^{*}\mu^{+}\tau^{-}$ (right).}
\label{fig:BRL}
\end{figure}
    \begin{figure}[h!]
\centering
\includegraphics[height=53mm,width=73mm]{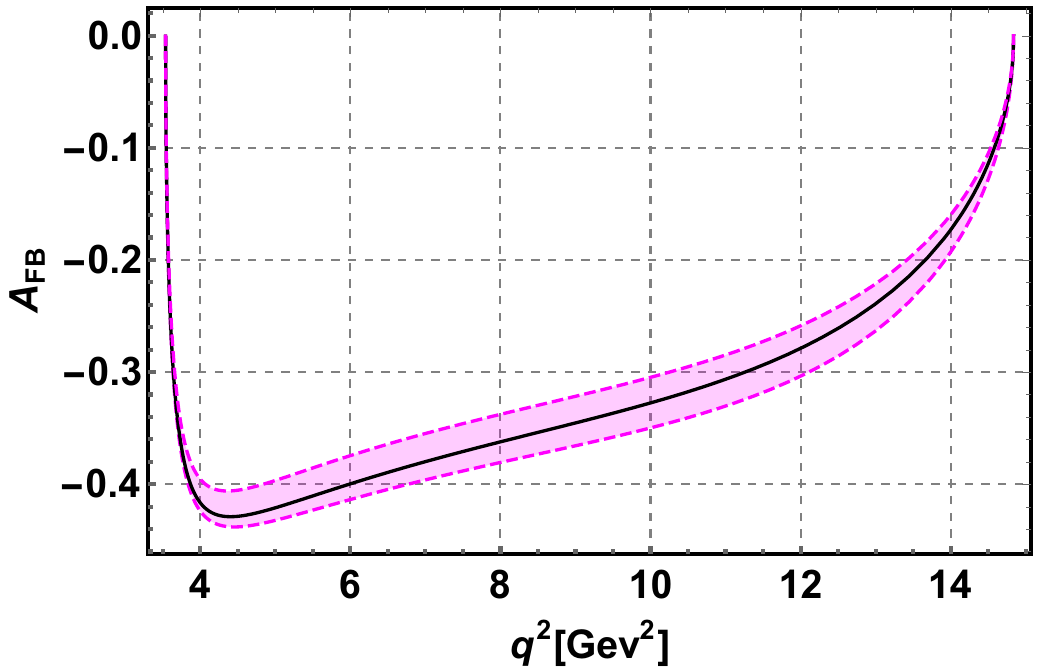}
\includegraphics[height=53mm,width=73mm]{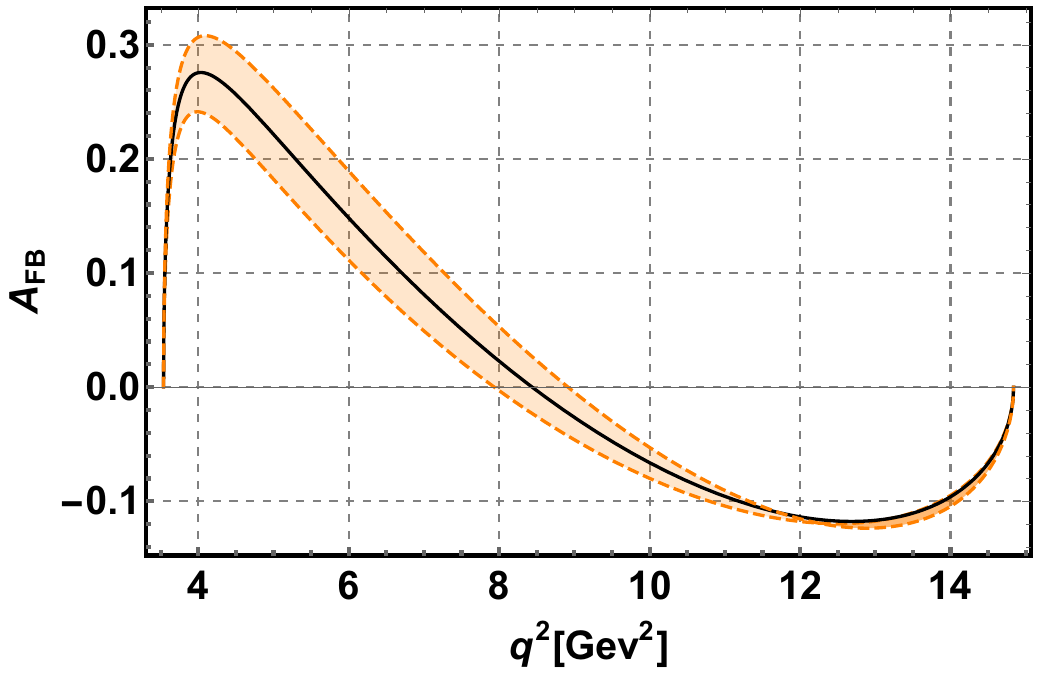}
\caption{Forward-backward asymmetry of the $B \to  K_{2}^{*}\tau^{+} \mu^{-}$ (left) and  $B \to K_{2}^{*} \tau^{-} \mu^{+}$(right).}
\label{fig:AFk2}
\end{figure}
\item $\textbf{Forward-backward asymmetry}$: Fig. \ref{fig:AFk2} depicts the $q^{2}$ dependency of the forward-backward asymmetry. The two plots illustrate distinct behaviors for the two different decay modes. For $B \to K_{2}^{*}\mu^{-}\tau^{+}$ decay, the observable $A_{FB}$ is negative throughout the entire $q^{2}$ region, while for $B \to K_{2}^{*}\mu^{+}\tau^{-}$ decay, the forward-backward asymmetry curve have a zero crossing around 8.5 ${\rm GeV}^{2}$ region.

\end{itemize} 
The numerical estimation is given in Table \ref{tab::Results}, whereas, the  cut-off dependent branching fraction is depicted in Fig. (\ref{fig:cutoffbr11})
\subsection{Impact of SMEFT NP coefficients on $B \to (K^{*} ,\phi )\ell_{1} \ell_{2}$  decay observables}
\begin{itemize}
    \item $\textbf{Branching ratio}$: The numerical values for the expected branching ratios of the decay process are provided in Table \ref{tab::Results}. In Fig. \ref{fig::BRM}, we present the $q^{2}$-dependent differential branching ratio of $B_{(s)} \to (K^{*},\phi) \tau^{\pm}\mu^{\mp}$ decay processes. Each mode of $B_{(s)} \to ( K^{*},\phi)$ process has almost similar  $q^2$ dependencies of the differential branching ratio irrespective of the charge of the heavier lepton. Therefore, we have shown a single curve for both modes $\tau^{\pm}\mu^{\mp}$.
    \begin{figure}[h!]
\centering
\includegraphics[height=55mm,width=75mm]{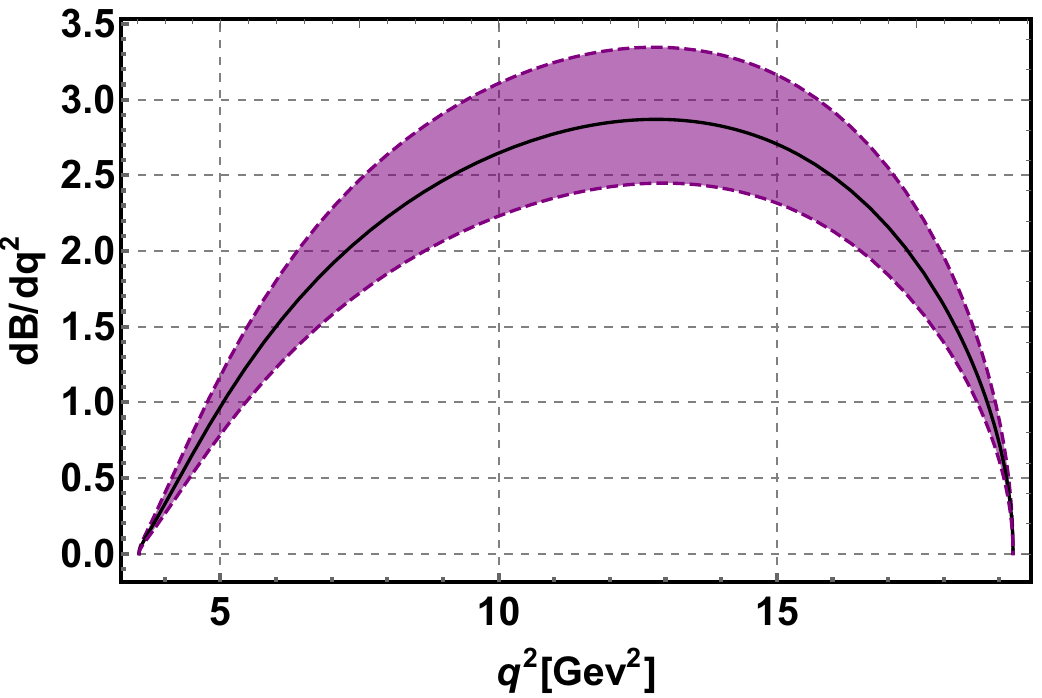}
\hspace{2mm}
\includegraphics[height=55mm,width=75mm]{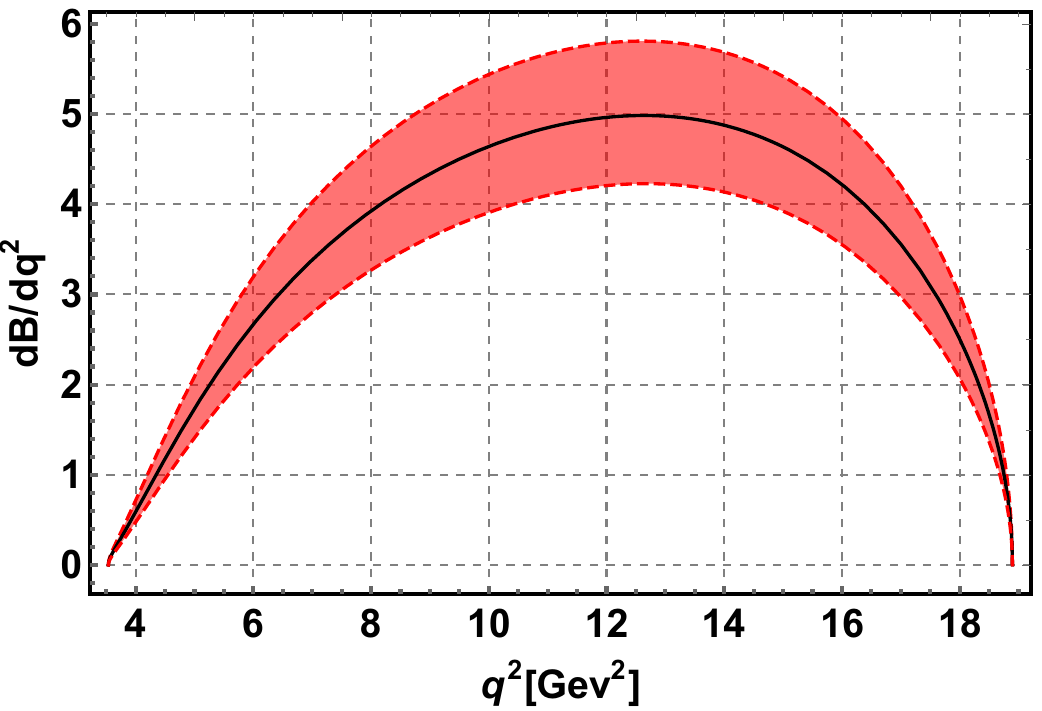}
\caption{Branching ratio (in the units of $10^{-6}$) of  $B \to  K^{*}\tau \mu $ (left) and  $B \to \phi \tau \mu $ (right).}
\label{fig::BRM}
\end{figure}
    \item $\textbf{Forward-backward asymmetry}$: We observe a zero-crossing position for both modes in the $\tau^- \mu^+$ final states, whereas no such position is found for $\mu^- \tau^+$ modes. The observable $A_{FB}$ exhibit the zero crossing around 7 GeV$^2$ for $B(B_{s}) \to K^* (\phi) \tau^- \mu^+$ processes. In contrast, for $B(B_{s}) \to K^* (\phi) \mu^- \tau^+$ channel, the behavior of the observable remains positive throughout the $q^2$ range. This is shown in Fig. \ref{fig::AFBM51} and Fig. \ref{fig::AFBM52}. We also provide the prediction for the forward-backward asymmetry for the decay process which is provided in Table \ref{tab::Results}.
    \begin{figure}[h!]
\centering
\includegraphics[height=55mm,width=75mm]{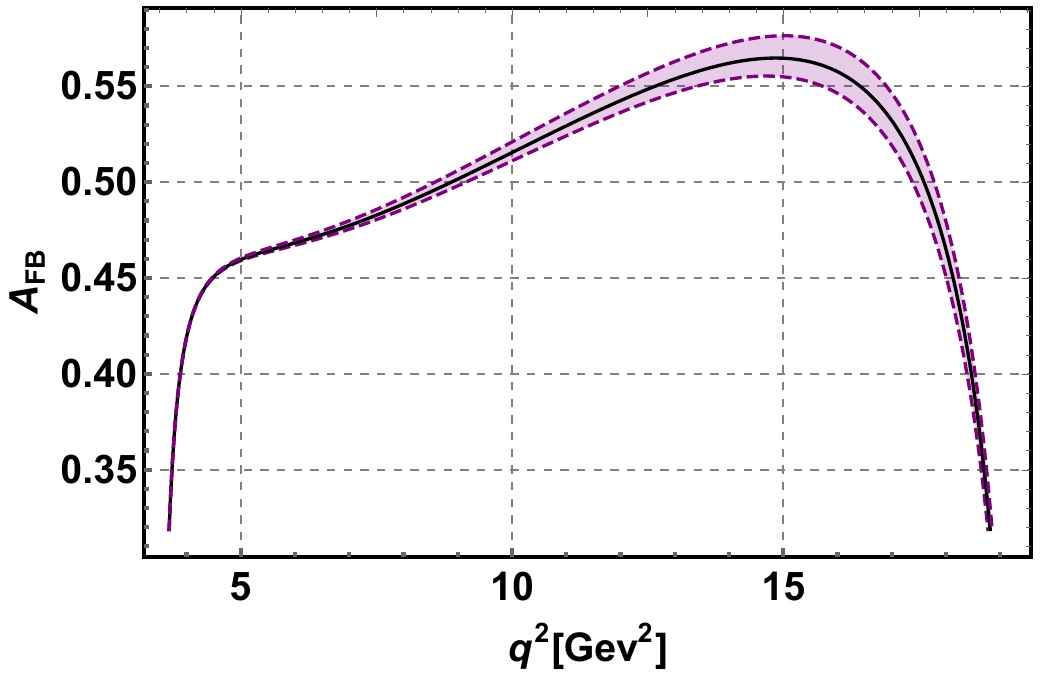}
\hspace{2mm}
\includegraphics[height=55mm,width=75mm]{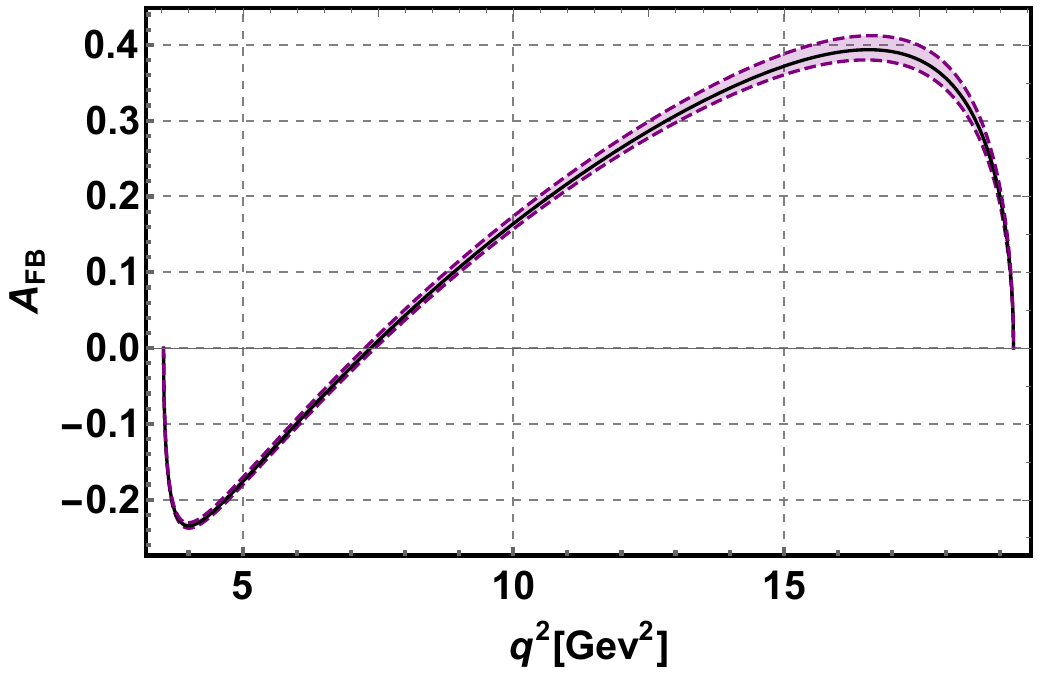}
\caption{Forward-backward asymmetry of $B \to  K^{*}\tau^{+} \mu^{-}$ (left) and  $B \to K^{*}\tau^{-} \mu^{+}$ (right)}
\label{fig::AFBM51}
\end{figure}
\begin{figure}[h!]
\centering
\includegraphics[height=55mm,width=75mm]{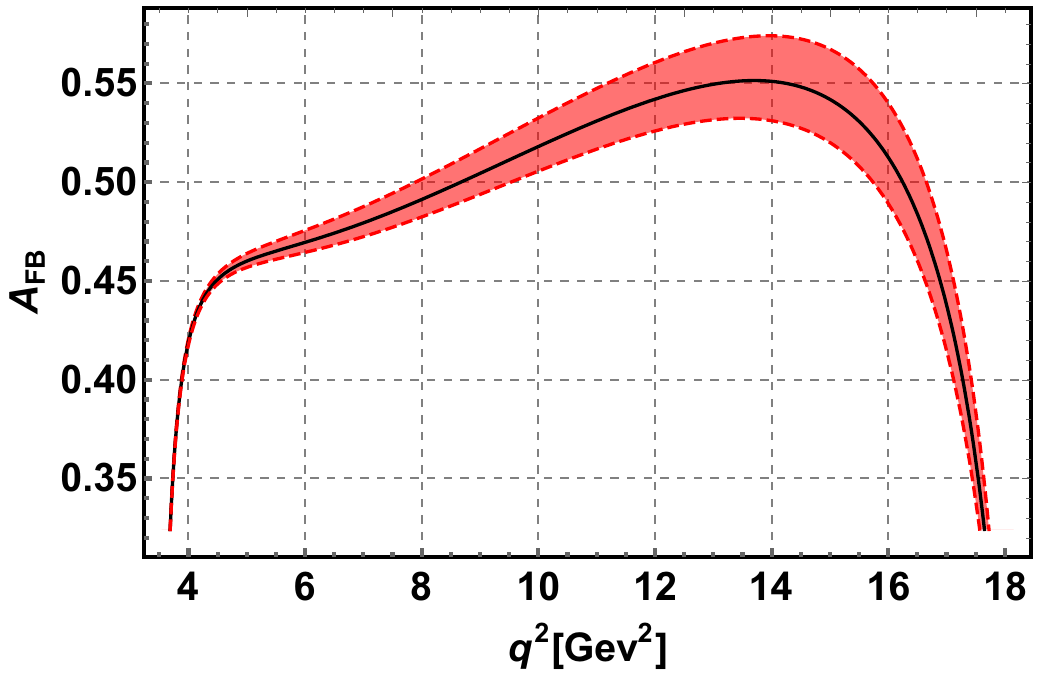}
\hspace{2mm}
\includegraphics[height=55mm,width=75mm]{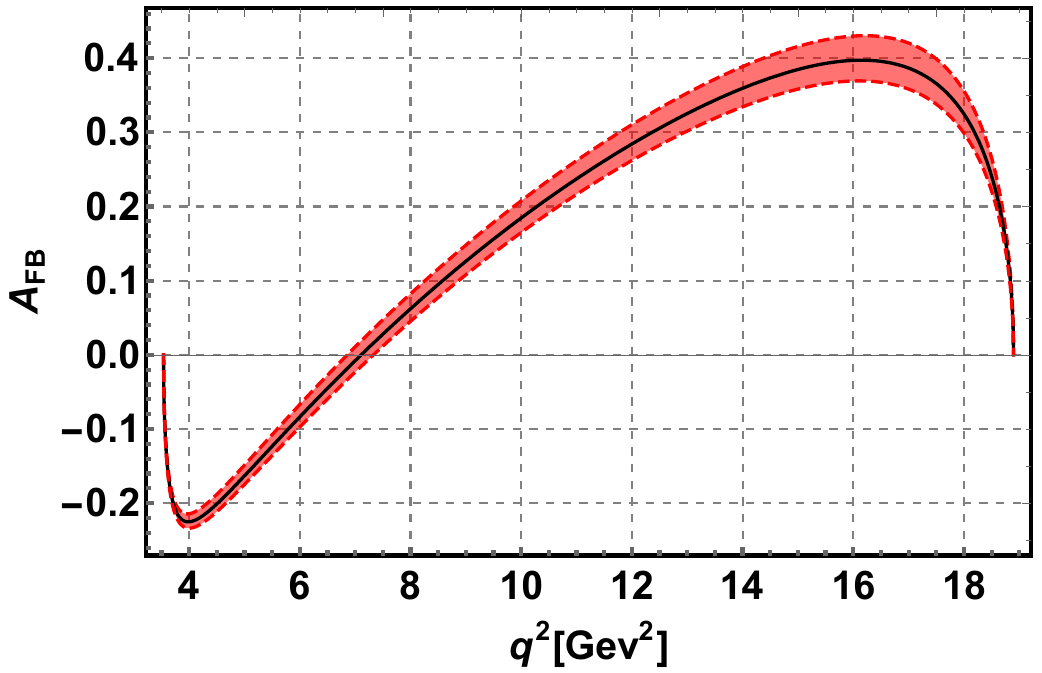}
\caption{Forward-backward asymmetry of $B_{s} \to \phi\tau^{+} \mu^{-}$ (left) and  $B_{s} \to \phi\tau^{-} \mu^{+}$ (right)}
\label{fig::AFBM52}
\end{figure}
\item $\textbf{Longitudinal polarisation asymmetry}$: The $q^2$ dependent lepton longitudinal polarisation asymmetry of $B \to K^{*}\tau \mu$ and $B \to \phi \tau \mu$ decays are shown in Fig. \ref{fig:FLM}. The observables have similar behavior irrespective of the charges of the lepton pair. In addition, the numerical values of $F_{L}$ are also given in the Table \ref{tab::Results}.
\begin{figure}[h!]
\centering
\includegraphics[height=55mm,width=75mm]{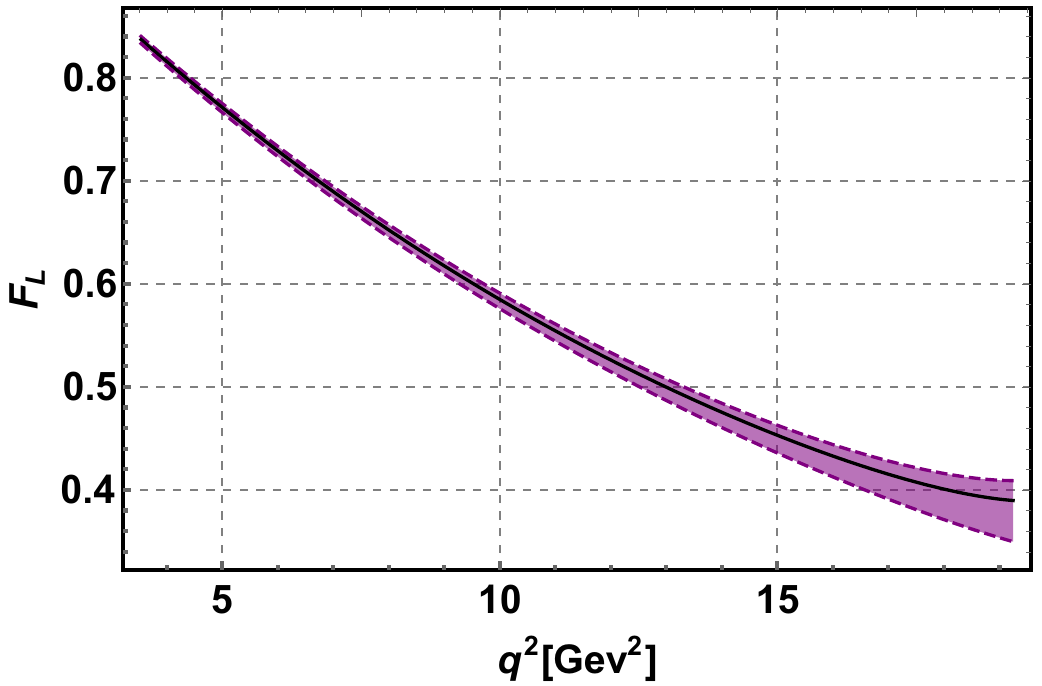}
\includegraphics[height=55mm,width=75mm]{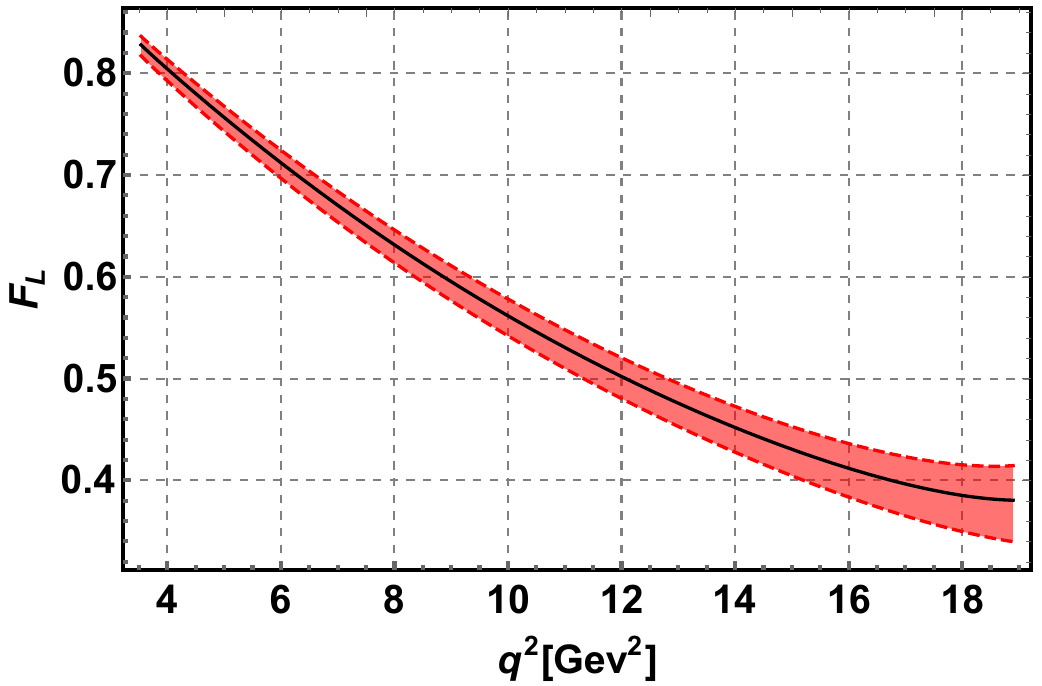}
\caption {Longitudinal polarization asymmetry of $B \to  K^{*}\tau \mu $ (left) and  $B \to \phi \tau  \mu$ (right)}
\label{fig:FLM}
\end{figure} 
\end{itemize}
\begin{figure}[h!]
\begin{center}
\includegraphics[height=55mm,width=75mm]{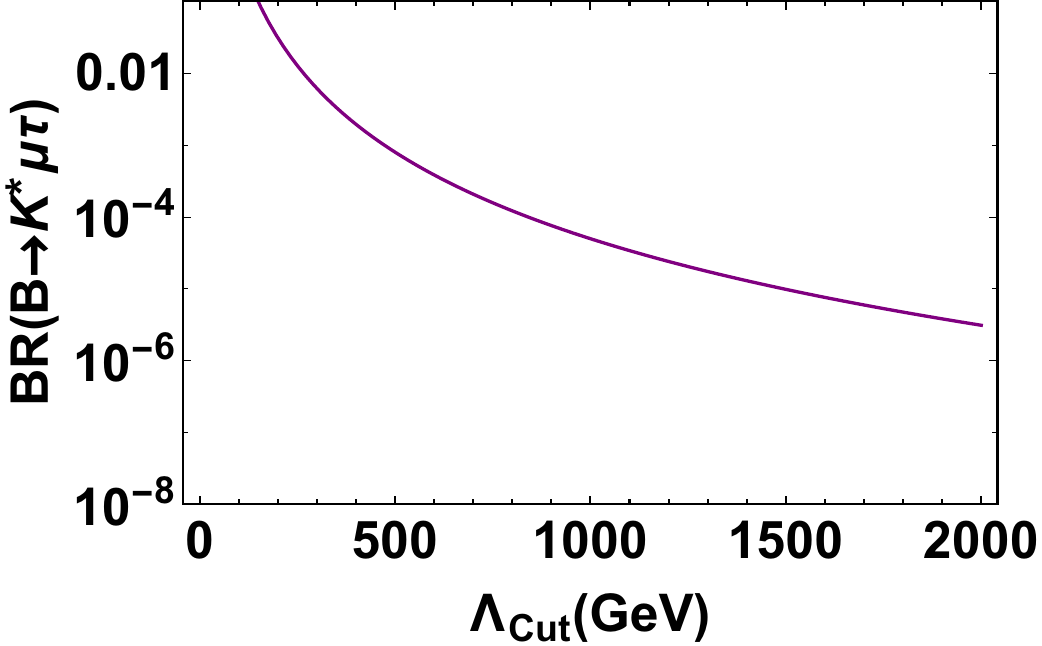}
\hspace{1mm}
\includegraphics[height=55mm,width=75mm]{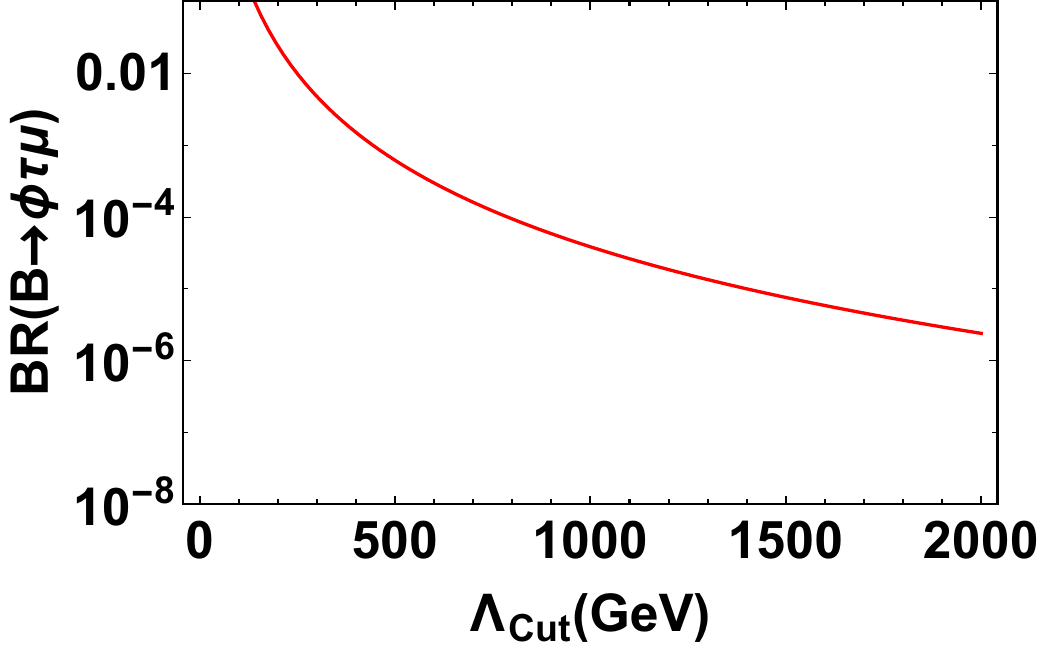}
\caption{ Cutoff dependent  branching ratio of $B \to  K^{*}\tau \mu $ (left) and $B \to \phi \tau \mu $ (right).}
\label{fig:cutoffbr}
\end{center}
\end{figure}  
As mentioned earlier, for our numerical computations, we set the cutoff limit $\Lambda_{\rm cut}$=1 TeV. The evolution of the branching ratio for the decay process is depicted in Fig. \ref{fig:cutoffbr}. 
\begin{table}[tbp]
\centering
\begin{tabular}{||c||c||c||c||}
\hline
\hline
\backslashbox{Decay mode}{Observable}
&\makebox[6em]{$\mathcal{B}$}&\makebox[10em]{$\mathcal{A}_{FB}$}&\makebox[10em]{$\mathcal{F}_L$}
\\\hline\hline
$\Lambda_{b}\to \Lambda \tau^{-}\mu^{+}$ & $\le 2.40\times 10^{-5}$ &$\le -0.294$ & $-$\\\hline 
$\Lambda_{b}\to \Lambda \tau^{+}\mu^{-}$ &$ \le 1.43\times 10^{-5}$ &$\le -0.053$ & $-$\\\hline  \hline
$B \to K^{*} \tau^{+} \mu^{-}$ & $\le 3.29 \times 10^{-5}$ &$\le 0.496$& $\le 0.565$\\ \hline
$B \to K^{*} \tau^{-} \mu^{+}$ & $\le3.26\times 10^{-5}$ &$\le0.165 $& $\le 0.561$\\ \hline \hline
$B_{s} \to \phi \tau^{-} \mu^{+}$ & $\le 5.61 \times 10^{-5}$ &$\le 0.174$& $\le 0.549$\\ \hline
$B_{s} \to \phi \tau^{+} \mu^{-}$ & $\le 5.66 \times 10^{-5}$ &$\le 0.496 $& $\le 0.552$\\ \hline \hline
$B \to K_{2}^{*} \tau^{+} \mu^{-}$ & $\le 4.71 \times 10^{-6}$ &$\le -0.320 $& $-$\\
\hline
$B \to K_{2}^{*} \tau^{-} \mu^{+}$ & $\le 4.38 \times 10^{-6}$ &$\le 0.016 $& $-$\\ \hline
\hline
\end{tabular} 
\caption{Predicted upper limits of $\mathcal{B}$ and $\mathcal{A_{FB}}$ for the $\Lambda_b\to\Lambda \tau^{\pm}\mu^{\mp}$, $B \to (K^{*}, \phi) \tau^{\pm} \mu^{\mp}$ and $B \to K_{2}^{*} \tau^{\pm} \mu^{\mp}$ decays, with $\mathcal{F_L}$ prediction for $B \to (K^{*}, \phi ) \tau^{\pm} \mu^{\mp}$.}
\label{tab::Results}
\end{table}

The above results are shown in the context of LFV $b$-hadron decays via the relevant WCs $C_{lq}^{(1)} $and $C_{leqd}$, considering the future experimental limits of $\mathcal{B}(B^{+}\to K^{+}\tau^{\pm}\mu^{\mp})$ and $\mathcal{B}(B_{s}\to \tau^{\pm}\mu^{\mp})$ decays. The essence for finding these numbers in  Table \ref{tab::Results} is that the BRs of these processes which are also induced by the mentioned WCs, should lie below their future limits. Therefore focusing on these numbers is crucial for the experimental point of view.

The decay modes discussed in our work so far have not been observed and only the experimental upper limit exists for $B \to K^{*} \tau\mu$ decay. In scenarios involving $C_{lq}^{(1)} $ and $C_{leqd}$, the upper limit for $\mathcal{B}(B \to K^{*} \tau\mu)$ ($\sim 10^{-5}$) aligns with the anticipated direct experimental upper limit \cite{Descotes-Genon:2023pen}. Similarly, for $B_{s} \to \phi \tau\mu$ decay, no direct experimental upper limit does exist. In our work, the upper limit for  BR is estimated to be ${\cal O}(10^{-5})$. It is interesting to note that these numbers fall within the scope of the future bounds that are expected to be obtained through upgrades at LHC \cite{Descotes-Genon:2023pen}. Taking into account for $B \to K_{2}^{*} \tau\mu$ and $\Lambda_{b}\to \Lambda \tau\mu$ processes, the estimated BRs are of the order $10^{-6}$ and $10^{-5}$, respectively. Although these two decay modes do not currently have any future bounds, it is anticipated that expected bounds could be obtained in upgrades at the LHC, similar to the previous two decays.
\section{Conclusion}
 The observation of any LFV decays indicates a clear signal of the presence of new physics beyond the SM, as such decays are strictly forbidden within its framework. Various extensions of the Standard Model have been proposed to explain lepton flavor-violating processes, which, in turn, motivate a closer examination of these decays. In this work, we have studied the lepton flavor violating $B_{(s)}$ decays based on $b \to s \tau \mu$ transitions. For this purpose, we constrained the parameter space of NP couplings weighted by (pseudo)scalar and axial(vector) operators,  by using the the upper limits of Br($B \to \tau \mu$) and Br($B \to K \tau \mu$) in the presence of SMEFT Wilson coefficients. Using such constrained NP parameters, we have investigated the impact on the branching fraction, forward-backward asymmetry and the lepton polarisation asymmetry of the exclusive $ B_{(s)} \rightarrow (\phi, K^{*}, K_{2}^{*})\tau\mu$  and  $\Lambda_{b}\rightarrow \Lambda \tau \mu$ processes. 
In the study of $\Lambda_{b}\to \Lambda  \tau \mu $ processes, the observables such as the branching fraction and the forward-backward asymmetries are discussed in our analysis. We observed that in the presence of new SMEFT couplings, these observables receive significant contributions. In the presence of NP coupling, the branching ratio for $\Lambda_{b}\to \Lambda  \tau^{-} \mu^{+} $ and $\Lambda_{b}\to \Lambda  \tau^{+} \mu^{-}$ decays are estimated to be $ 1.43\times 10^{-5}$ and $ 2.40\times 10^{-5}$, respectively. In contrast, the zero-crossing in the forward-backward asymmetry curve was found only for the $\Lambda_b \to \Lambda \tau^+ \mu^-$ mode. For the decays $B \to K^* \tau^- \mu^+$ and $B \to K^* \tau^+ \mu^-$, the branching ratios are estimated to be of the order of $10^{-5}$. Similarly, for $B_s \to \phi \tau^- \mu^+$ and $B_s \to \phi \tau^+ \mu^-$ decays, the predicted branching ratios are also of $\mathcal{O}(10^{-5})$, with minimal difference in central values. However, in this context, the branching fraction for the $B \to K_2^* \tau^{\pm} \mu^{\mp}$ decay is predicted to be of $\mathcal{O}(10^{-6})$. Furthermore, the zero-crossing point was obtained for the $B \to (K_2^*, \phi, K^*) \mu^+ \tau^-$ modes within this framework.

 Furthermore, we presented the cut-off scale dependency of the branching ratios for the $b \to s \tau \mu$ decays. Additionally, we made predictions for the upper limits of the aforementioned observables as summarized in Table \ref{tab::Results}. The findings reveal that our predicted values are substantial and within the reach of current or forthcoming experimental limits. If detected in future experiments, this would offer a definitive indication of new physics. 

\section{Acknowledgement}
DP would like to acknowledge the support of the Prime Minister's Research Fellowship, Government of India. MKM acknowledges IoE PDRF, University of Hyderabad for the financial support. RM would like to acknowledge the University of Hyderabad IoE project grant no. RC1-20-012.


\appendix 
\section{The Angular coefficients of $\Lambda_{b} \to \Lambda \ell_{1} \ell_{2}$ process}
The of the angular coefficients $K_{1ss,1cc,1c}$ can be expressed as follows:
\begin{align}
K_{1ss,1cc}=K_{1ss,1cc}^{VA}+K_{1ss,1cc}^{SP}+K_{1ss,1cc}^{int},
\end{align} 
where, the $K_{1ss,1cc}^{VA}$, $K_{1ss,1cc}^{SP}$ and $K_{1ss,1cc}^{int}$ are given as follows,
\begin{eqnarray}
K_{1ss}^{\rm VA} &=& \frac{1}{4} \bigg( 2|A_{||0}^R|^2 + |A_{||1}^R|^2 +
2|A_{\perp0}^R|^2 + |A_{||1}^R|^2 + \{ R \leftrightarrow L  \} \bigg) \nn\\
&&- \frac{m_+^2+m_-^2}{4q^2} \bigg[ \bigg( |A^R_{\|_0}|^2 + |A^R_{\perp_0}|^2 + \{ R \leftrightarrow L \} \bigg) - \bigg( |A_{\perp t}|^2  + \{ \perp \leftrightarrow \| \}\bigg) \bigg] \nn\\
&&+ \frac{m_+^2-m_-^2}{4q^2} \bigg[ 2 \rm Re \bigg( A^R_{\perp_0} A^{\ast L}_{\perp_0} + A^R_{\perp_1}A^{\ast L}_{\perp_1} + \{ \perp \leftrightarrow \| \}  \bigg) \bigg] \nn\\
&&- \frac{m_+^2m_-^2}{4q^4} \bigg[ \bigg(|A_{\perp 1}^R|^2 + |A_{||1}^R|^2 
+ \{ R \leftrightarrow L \} \bigg) + 2|A_{\|t}|^2 + 2|A_{\perp t}|^2\bigg] , 
\end{eqnarray}
\begin{eqnarray}
K_{1cc}^{\rm VA} &=& \frac{1}{2}\bigg( |A_{\perp 1}^R|^2 + |A_{||1}^R|^2 + \{R \leftrightarrow L \} \bigg) + \frac{m_+^2+m_-^2}{4q^2} \nn\\
&& \times \bigg[ \bigg (|A^R_{||_0}|^2 - |A^R_{||_1}|^2 + |A^R_{\perp_0}|^2 - |A^R_{\perp_1}|^2 + \{ R \leftrightarrow L \} \bigg) + \bigg( |A_{\perp t}|^2 + |A_{|| t}|^2 \bigg) \bigg] \nn\\ 
&&+ \frac{m_+^2-m_-^2}{4q^2} \bigg[ 2 \rm Re \bigg( A^R_{\perp_0}A^{* L}_{\perp_0}\ + A^R_{\perp_1}A^ L_{\perp _1} + \{\perp \leftrightarrow || \} \bigg) \bigg] \nn\\
&&- \frac{m_+^2 m_-^2}{2q^4}  \bigg[ \bigg(|A_{\perp0}^{R}|^2 + |A_{||0}^{R}|^2 + \{ R \leftrightarrow L \} \bigg) + |A_{||t}|^2 + |A_{\perp t}|^2\bigg] ,\\
K_{1c}^{\rm VA} &=& -\beta_\ell \beta_\ell^ \prime \bigg( A^R_{\perp_1} A^{* R}_{||_1} - \{ R \leftrightarrow L \}  \bigg) + \beta_\ell \beta_\ell^ \prime \frac{m_+ m_-}{q^2} \rm Re \bigg( A_{||0}^L A_{|| t}^ * + A_{\perp 0}^ L A_{\perp t}^* \bigg).
\end{eqnarray}
\begin{align}
&K_{1ss}^{\rm SP} = \frac{1}{4}\bigg( |A_{\rm S\perp}|^2 + |A_{\rm P \perp}|^2 + \{ \perp \leftrightarrow \| \} \bigg) - \frac{m_+^2}{4q^2}\big(|A_{S ||}|^2 + |A_{S \perp}|^2\big) - \frac{m_-^2}{4q^2}\big(|A_{P ||}|^2 + |A_{P \perp}|^2\big), \nn\\
& K_{1cc}^{\rm SP} = \frac{1}{4}\bigg( |A_{\rm P\perp}|^{2} + |A_{\rm S\perp}|^{2} + \{\perp \leftrightarrow \| \} \bigg) - \frac{m_+^2}{4q^2}\big(|A_{S ||}|^2 + |A_{S \perp}|^{2}\big) - \frac{m_-^2}{4q^{2}}\big(|A_{P ||}|^{2} + |A_{P \perp}|^{2}\big), \nn\\
&K_{1c}^{\rm SP} = 0.
\end{align}
\begin{eqnarray}
K_{1ss}^{\rm int} &=& \frac{m_+}{2\sqrt{q^2}} \rm Re \bigg( A_{||t}A_{P||}^* + A_{\perp t}A_{P \perp}^*  \bigg) +
\frac{m_-}{2\sqrt{q^2}} Re \bigg( A_{||t}A_{S||}^* + A_{\perp t}A_{S \perp}^*  \bigg) \nn\\
&&- \frac{m_+^2m_-}{2q^2\sqrt{q^2}} Re \bigg( A_{||t}A_{S||}^* + A_{\perp t}A_{S \perp}^* \bigg) - \frac{m_+m_-^2}{2q^2\sqrt{q^2}} Re \bigg(A_{||t}A_{P||}^* + A_{\perp t}A_{P \perp}^* \bigg), \\
 K_{1cc}^{\rm int} &=& \frac{m_+}{2\sqrt{q^2}}Re \bigg(A_{||t}A_{P||}^* + A_{\perp t}A_{P\perp }^* \bigg) + \frac{m_-}{2\sqrt{q^2}}Re \bigg(A_{||t}A_{S||}^*
+ A_{\perp t}A_{S \perp}^* \bigg) \nn\\ 
&&- \frac{m_+^2m_-}{2q^2 \sqrt{q^2}} Re \bigg( A_{||t}A_{S||}^{*} + A_{\perp t}A_{S\perp}^{*} \bigg) - \frac{m_+ m_-^2}{2q^2\sqrt{q^2}}Re \bigg( A_{||t}A_{P||}^{*} + A_{\perp t}A_{P\perp}^{*} \bigg), \\
K_{1c}^{\rm int} &=& \frac{\beta_\ell\beta_\ell^\prime}{2\sqrt{q^2}} Re \bigg( A_{S||}A_{||0}^{L*} + A_{S\perp}A_{\perp 0}^{L*} + A_{S||}A_{||0}^{R*}+A_{|S\perp}A_{\perp 0}^{R*}  \bigg) \nn\\ 
&&+ \frac{\beta_\ell\beta_\ell^\prime}{2\sqrt{q^2}} Re \bigg( A_{P||}A_{||0}^{L*} + A_{P\perp}A_{\perp0}^{L*} +-A_{P||}A_{||0}^{R*}-A_{|S\perp}A_{\perp 0}^{R*} \bigg).
\end{eqnarray}  
Here, we have defined $ m_{\mp}=m_{1} \mp m_{2} $ with $m_1, m_2 $ as the masses of $\ell_1$ and $\ell_2$, respectively.
\section{Inputs required for $B \to K_2^* \ell _1 \ell _2$ decay mode} \label{inputsBtoK2star}
\begin{eqnarray}
C(q^2)&=&\frac{3}{8}\beta_{+}^2\beta_{-}^2   \left\lbrace\left(|A_L^{\parallel}|^2+|A_L^{\perp}|^2-2|A_L^{0}|^2\right)+\left(L\to R\right)\right\rbrace,\\ \label{dist:C}
B(q^2)&=&\frac{3}{2}\beta_+\beta_-\left \lbrace \rm Re \left[A_L^{\perp *} A_L ^\parallel -(L \to R)\right]+ \frac{m_+m_-}{q^2} \rm Re\left[A_L^{0*} A_L^{t} +(L \to R)\right]\right.\nn\\
&& \left. + \frac{m_+}{\sqrt{q^2}} \rm Re\left[A_S^* (A_L^0 +A_R^0) \right]-\frac{m_-} {\sqrt{q^2}}\rm Re \left[A_{SP}^* (A_L^0 -A_R^0) \right]\right\rbrace,\\
A(q^2) &=& \frac{3}{4}\left\lbrace\frac{1}{4}\left[\left(1+\frac{m_{+}^{2}}{q^2}\right)\beta_{-}^{2}+\left(1+\frac{m_{-}^{2}}{q^2}\right)\beta_{+}^2 \right]\left(|A_L^{\parallel}|^2+|A_L^{\perp}|^2+(L\to R)\right)\right.\nn\\
&&+\frac{1}{2}\left(\beta_{-}^2 +\beta_{+}^2\right)\left(|A_L^{0}|^2+|A_R^{0}|^2\right)+\frac{4m_1 m_2}{q^2} \rm Re\left[ A_L^{0*} A_R^0 + A_L^{\parallel *}A_R^{\parallel}+A_R^{\perp} A_L^{\perp *}-A_L^t A_R^{t*}\right] \nn\\
&&+\frac{1}{2}\left(\beta_+^2+\beta_-^2-2\beta_+^2\beta_-^2\right)\left(|A_L^t|^2 + |A_R^t|^2\right) +\frac{1}{2} \left(|A_{SP}|^{22} +|A_{S}|^{22}\right) \nn\\
&& +\frac{2m_-}{\sqrt{q^2}}\beta_+^2 \rm Re\left[A_S(A_L^t+A_R^t)^* \right]- \frac{2m_+}{\sqrt{q^2}} \beta_-^2 \rm Re\left[A_{SP}(A_L^t- A_R^t)^*]\right \rbrace\label{dist:A}.  
 \end{eqnarray}
\section{Inputs required for $B \to (K^*, \phi) \ell _1 \ell _2$ process} \label{InputsBtoV}
The full angular distribution for the decay process can  be expressed as
\begin{equation}
\begin{split}
\frac{d^4\Gamma(B\rightarrow\bar{K}^*(\rightarrow K\pi)\ell_{\alpha}^-\ell_{\beta}^+)}{dq^2dcos\theta_{\ell}dcos\theta_{K}d\phi}=\frac{9}{32\pi}I(q^2,\theta_{\ell},\theta_K,\phi),
\end{split}
\end{equation}
where \\
\begin{align}
I(q^2,\theta_\ell,\theta_K,\phi) = & I_1^s(q^2)\sin^2\theta_K+I_1^c(q^2)\cos^2\theta_K +[I_2^s(q^2)\sin^2\theta_K+I_2^c(q^2)\cos^2\theta_K]\cos 2\theta_\ell\nn\\[.4em] 
&+I_3(q^2)\sin^2\theta_K \sin^2\theta_\ell \cos 2\phi+I_4(q^2)\sin 2\theta_K \sin 2\theta_\ell \cos \phi \nn \\[.4em] 
&+ I_5(q^2) \sin 2\theta_K\sin \theta_\ell\cos\phi+[I_6^s(q^2)\sin^2\theta_K+I_6^c(q^2)\cos^2\theta_K]\cos \theta_\ell \nonumber \\[.4em] 
&+I_7(q^2)\sin 2\theta_K \sin \theta_\ell \sin \phi + I_8(q^2)\sin 2\theta_K \sin 2\theta_\ell \sin\phi \nonumber\\[.4em] 
&+I_9(q^2) \sin^2\theta_K \sin^2\theta_\ell \sin 2 \phi,
\end{align}
with $\theta_{\ell}$ , $\theta_{K}$ and $\phi$ varies as $-\pi \le \theta_{\ell}, \theta_{K}\le \pi$ and $0\le \phi \le 2\pi$ respectively. \\
The angular coefficients in terms of transversity amplitude can be expressed as:
\begin{align}
\label{eq:angular}
I_1^s(q^2) &=\frac{4 m_1 m_2}{q^2}\mathrm{Re}\left(A_{\parallel}^L A_{\parallel}^{R\ast}+A_{\perp}^L A_{\perp}^{R\ast}\right)+\biggl[|A_{\perp}^L|^2+|A_{\parallel}^L|^2+ (L\to R) \biggr]\frac{\lambda_q +2 [q^4-(m_1^2-m_2^2)^2]}{4 q^4}, \nn \\
I_1^c(q^2) &= \frac{8 m_1 m_2}{q^2} \mathrm{Re}(A_0^L A_0^{R\ast}-A_t^L A_t^{R\ast}) +\bigl[|A_0^L|^2+|A_0^R|^2 \bigr]\frac{q^4-(m_1^2-m_2^2)^2}{q^4}\nn\\
&\hspace{3.5cm}-2\frac{(m_1^2-m_2^2)^2-q^2 (m_1^2+m_2^2)}{q^4}\bigl(|A_t^L|^2+|A_t^R|^2\bigr),\nonumber\\
I_2^s(q^2) &= \frac{\lambda_q}{4 q^4}[|A_\parallel^L|^2+|A_\perp^L|^2+(L\to R)], \nn  \\
I_2^c(q^2) &= - \frac{\lambda_q}{q^4}(|A_0^R|^2+|A_0^L|^2), \nn \\
I_3(q^2) &= \frac{\lambda_q}{2 q^4} [|A_\perp^L|^2-|A_\parallel^L|^2+(L\to R)],\nn \\
I_4(q^2) &= - \frac{\lambda_q}{\sqrt{2} q^4} \mathrm{Re}(A_\parallel^L A_0^{L\ast}+(L\to R)],\nn \\
I_5(q^2) &= \frac{\sqrt{2}\lambda_q^{1/2}}{q^2} \left[ \mathrm{Re}(A_0^L A_\perp^{L\ast}-(L\to R)) -\frac{m_1^2-m_2^2}{q^2} \mathrm{Re}(A_t^L A_\parallel^{L\ast}+(L\to R))\right], \nn \\
I_6^s(q^2) &=- \frac{2 \lambda_q^{1/2}}{q^2}[\mathrm{Re}(A_\parallel^L A_\perp^{L\ast}-(L\to R))],\nn  \\
I_6^c(q^2) &= - \frac{4\lambda_q^{1/2}}{q^2}\frac{m_1^2-m_2^2}{q^2} \mathrm{Re}(A_0^L A_t^{L\ast}+(L\to R)),\nn \\
I_7(q^2) &= - \frac{\sqrt{2}\lambda_q^{1/2}}{q^2} \left[ \frac{m_1^2-m_2^2}{q^2} \mathrm{Im}(A_\perp^{L}A_t^{L\ast} +(L\to R))+ \mathrm{Im}(A_0^L A_\parallel^{L\ast}-(L\to R))\right], \nn \\
I_8(q^2) &= \frac{\lambda_q}{\sqrt{2}q^4}\mathrm{Im}(A_0^{L}A_\perp^{L\ast} +(L\to R)), \nn \\
I_9(q^2) &=- \frac{\lambda_q}{q^4}\mathrm{Im}(A_\perp^R A_\parallel^{R\ast}+A_\perp^L A_\parallel^{L\ast}   ).
\end{align}
The transversity amplitude can be written as follows \cite{Becirevic:2016zri},
\begin{align}
A_{\perp}^{L(R)} =& {\cal N}_{K^\ast} \sqrt{2} \lambda_B^{1/2}\left[(C_9 \mp C_{10} )\frac{V(q^2)}{m_B+m_{K^\ast}}\right],\nn\\
\hspace{2mm}
A_{\parallel}^{L(R)} =&  -{\cal N}_{K^\ast} \sqrt{2}(m_B^2-m_{K^\ast}^2)\left[(C_9 \mp C_{10})\frac{A_1(q^2)}{m_B-m_{K^\ast}}\right],\nn\\
A_0^{L(R)}=&-\frac{{\cal N}_{K^\ast}}{2 m_{K^\ast} \sqrt{q^2}}(C_9 \mp C_{10})\left[ (m_B^2-m_{K^\ast}^2-q^2)(m_B+m_{K^\ast})A_1(q^2)-\frac{ \lambda_B A_2(q^2)}{m_B+m_{K^\ast}}\right],  \nonumber\\[0.7 em]
A_{t}^{L(R)} =&  -{\cal N}_{K^\ast} \frac{\lambda_B^{1/2}}{\sqrt{q^2}}\left[(C_9  \mp C_{10}) +\frac{q^2}{m_b+m_s}\left(\frac{C_S}{m_1-m_2}\mp \frac{C_P}{m_1+m_2}\right)\right] A_0(q^2).
\end{align} \label{eq:helicityamplitudest}\\

where,

\begin{align}
\label{eq:NK}
 N_{K^*}=&V_{tb}V_{ts}^*\bigg[\frac{\tau_{B_d}G_{F}^2\alpha_{em}^2 \sqrt{\lambda_{B}\lambda_q}}{3\times2^{10}\pi^{5}m_{B}^3}\bigg]^{\frac{1}{2}  },\nn\\
 \lambda(a^2 ,b^2 ,c^2)=& a^4 + b^4 +c^4 -2(a^2b^2 +b^2c^2+c^2a^2).
\end{align}
In the above equation, $G_F$ is the Fermi coupling constant whereas $\tau_{B_d}$ represents the life time of $B_d$ meson. The term proportional to $V_{tb} V_{ts}$ survives in the SM limit. 
\bibliographystyle{ieeetr}
\bibliography{RDM}

\end{document}